\documentclass[prd,showpacs,amsmath,amssymb]{revtex4}
\usepackage[dvips]{graphicx}

\begin{document}
\title{Relation between the Polyakov loop and the chiral order
       parameter at strong coupling}
\author{Kenji Fukushima}
\email[Electronic address: ]{fuku@nt.phys.s.u-tokyo.ac.jp}
\affiliation{Department of Physics, University of Tokyo, 7-3-1 Hongo,
             Bunkyo-ku, Tokyo 113-0033, Japan}
\begin{abstract}
We discuss the relation between the Polyakov loop and the chiral order
parameter at finite temperature by using the Gocksch-Ogilvie model
with fundamental or adjoint quarks. The model is based on the double
expansion of strong coupling and large dimensionality on the lattice.
In an analytic way with the mean field approximation employed, we show
that the confined phase must be accompanied by the spontaneous
breaking of the chiral symmetry for both fundamental and adjoint
quarks. Then we proceed to numerical analysis to look into
the coupled dynamics of the Polyakov loop and the chiral order
parameter. In the case of fundamental quarks, the pseudo-critical
temperature inferred from the Polyakov loop behavior turns out to
coincide with the pseudo-critical temperature of the chiral phase
transition. We discuss the physical implication of the coincidence of
the pseudo-critical temperatures in two extreme cases; one is the
deconfinement dominance and the other is the chiral dominance. As for
adjoint quarks, the deconfinement transition of first order persists
and the chiral phase transition occurs distinctly at higher
temperature than the deconfinement transition does. The present model
study gives us a plausible picture to understand the results from the
lattice QCD and aQCD simulations.
\end{abstract}
\maketitle


\section{INTRODUCTION}

Recent developments in the heavy ion collision experiment are going to
enable us to reach phase transitions that the early universe must have
undergone, i.e., the phase transition from hadronic matter to a
quark-gluon plasma (QGP) \cite{qgp}. From the theoretical point of
view what is known as the QGP phase transition can consist of two
distinct phase transitions. One is the \textit{color deconfinement
transition} and the other is the \textit{chiral symmetry restoration}
or the \textit{chiral phase transition}. The color liberation
associated with deconfinement is a phenomenon distinguishable from the
chiral phase transition in principle, but it is likely that the
mechanism of confinement would be closely related to the chiral
dynamics. According to several theoretical studies \cite{cas79}, in
fact, confinement should be accompanied by the spontaneous chiral
symmetry breaking. In other words, the critical temperature relevant
to the deconfinement transition (denoted by $T_{\text{d}}$) is
inevitably lower than that relevant to the chiral phase transition
(denoted by $T_\chi$), that is; $T_{\text{d}}\le T_\chi$, if two phase
transitions are to be distinguished.

Actually it is difficult to shed light on underlying physics in the
relation between two phase transitions. It is partly because most of
the effective models involving the chiral dynamics have shortcomings
in describing confinement physics, and partly because the criterion to
characterize the deconfinement transition cannot go together with the
realization of the chiral symmetry. We can define an order parameter
for the deconfinement transition only when there is no dynamical quark
in a theory, or quarks are infinitely massive ($m_q\to\infty$). In
contrast to it, the chiral symmetry is manifested when quarks can be
regarded as massless ($m_q\sim0$). Thus two phase transitions must lie
in the opposite limits in regard to the current quark mass.

In the absence of dynamical quarks pure gluonic systems have the
center symmetry and the Polyakov loop, $L$, provides a criterion for
confinement \cite{pol78,mcl81}. The expectation value of the Polyakov
loop traced in the fundamental representation can be related to the
free energy $f_q$ with a static fundamental quark placed in a thermal
medium, that is; $\langle\mathrm{Tr_c}L\rangle=\exp[-\beta f_q]$,
where $\mathrm{Tr_c}$ denotes the trace over the color space and
$\beta$ is the inverse temperature \footnote{We would like to comment
upon the subtleties encountered in defining and interpreting the
Polyakov loop in a continuum theory. Unless the color group is SU(2),
$\langle\mathrm{Tr_c}L\rangle$ may be complex, while $\exp[-\beta f_q]$
is always real. Moreover it is known that
$\langle\mathrm{Tr_c}L\rangle$ without any proper renormalization goes
to zero in the continuum limit. As emphasized in Ref.~\cite{smi94}
what is meaningful is the correlation function of the Polyakov loops,
from which the inter-quark potential can be inferred. Although the
Polyakov loop might make little sense as it is, however, the properly
renormalized Polyakov loop can be defined by the correlation function
of the Polyakov loops with infinite separation. Recently it has been
actually demonstrated that the renormalized Polyakov loop is
well-behaved as an order parameter \cite{kac02}. We will not go into
this issue in the present paper since we take an ansatz for the mean
field approximation (the Polyakov loop is then chosen as real) and
have an explicit cut-off parameter of the lattice spacing here.}.
Under the center transformation the traced Polyakov loop,
$\mathrm{Tr_c}L$, is twisted by an element of the center group (i.e.\
the Z($N_{\text{c}}$) group for an SU($N_{\text{c}}$) gauge
theory). In the confined phase where $f_q=\infty$, the center symmetry
is realized to result in $\langle\mathrm{Tr_c}L\rangle=0$, while the
spontaneous breaking of the center symmetry occurs in the deconfined
phase where the expectation value of the Polyakov loop takes a finite
value corresponding to $f_q<\infty$.

For the purpose of looking into the spontaneous breaking of the center
symmetry, the effective action in terms of the Polyakov loop has been
pursued by the strong coupling expansion as well as by the
perturbative calculation. Indeed, it has been revealed that the center
symmetry is spontaneously broken in a perturbative regime at high
temperature, which is actually embodied in what is often called the
Weiss potential \cite{wei81}. As the temperature is lowered, unstable
modes can develop to make the expectation value of the Polyakov loop
diminished \cite{fuk00}. Unfortunately, however, we cannot reach an
appropriate description of the deconfinement transition in any
perturbative way because the phase transition is a non-perturbative
phenomenon and moreover the perturbative calculation would have no
information on the Haar measure \cite{fuk00,goc93}. The Haar measure
in the context of confinement physics means a sort of the
Faddeev-Popov determinant \cite{rei96}, or precisely speaking, the
Jacobian of the variable transformation from the gauge potential to
the Polyakov loop. As stressed also in Ref.~\cite{len98}, the nature
of the group integration contained in the Haar measure can be
responsible for confinement and, in fact, it is the case in the strong
coupling expansion on the lattice. It has been found that the
resultant effective action in the lowest order of the strong coupling
expansion leads to a second order phase transition for the SU(2) gauge
theory \cite{pol82} and a first order phase transition for the
SU($N_{\text{c}}\ge3$) gauge theories \cite{gro83}. These results in
the strong coupling expansion are consistent with the anticipation
derived from the analogy to spin systems based on the center symmetry
\cite{pol78}. They are in agreement with the results from the lattice
simulation as well \cite{mcl81}.

When massless quarks in the fundamental representation are present in
a theory, the center symmetry is broken explicitly and the Polyakov
loop would no longer serve as an order parameter to identify the
deconfinement transition \cite{ban83}. Then, another symmetry, i.e.,
the chiral symmetry plays an important role in hadronic
properties. There are many effective approaches based on the chiral
symmetry such as the linear-sigma model, the Nambu--Jona-Lasinio
model, the chiral random matrix model and so on \cite{hat94}. The
spontaneous breaking of the chiral symmetry with the staggered fermion
on the lattice has already been investigated by means of the double
expansion of strong coupling and large dimensionality
\cite{klu83,dam86}. The results are surprisingly acceptable not only
qualitatively but quantitatively also.

For a moderate value of the quark mass, neither the deconfinement
transition nor the chiral phase transition has any well-defined order
parameter. In the lattice QCD simulation, it has been observed that
the Polyakov loop and the chiral order parameter show crossover
behaviors in the presence of dynamical quarks \cite{kar94}, as if the
magnetization in spin systems under an external magnetic field. To be
interesting, the lattice QCD results suggest that the simultaneous
transitions for deconfinement and chiral restoration take place at the
same pseudo-critical temperature, that is; the respective peaks of the
susceptibility for the Polyakov loop and for the chiral order
parameter are found at the same temperature.

In Ref.~\cite{sat98}, Satz emphasized that such coincidence of the
pseudo-critical temperatures can be understood in the following way:
At low temperature where the chiral symmetry is spontaneously broken,
the explicit breaking of the center symmetry is suppressed by the
heavier mass of constituent quarks rather than the lighter mass of
current quarks. Consequently the expectation value of the Polyakov
loop stays small in the confined or chiral broken phase at low
temperature. Once the constituent quark mass drops off in the chiral
symmetric phase at high temperature, the explicit breaking of the
center symmetry is enlarged to result in a finite value of the
Polyakov loop. Although this qualitative argument is developed further
in Ref.~\cite{dig01}, any quantitative demonstration had not been
given until our analysis on the Gocksch-Ogilvie model
\cite{goc85,fuk03}. It should be noticed that, according to Satz's
argument, the behavior of the Polyakov loop would be sensitive to the
chiral phase transition rather than the deconfinement transition. In
order to disclose the underlying relation between deconfinement and
chiral restoration, some legitimate order parameters are indispensable
to characterize the deconfinement transition even in the presence of
dynamical quarks. However, defining an order parameter for the
deconfinement transition is a quite difficult question because of the
thermal excitation of dynamical quarks, particularly under the
thermodynamic limit at finite temperature \cite{fuk03_2}.

Alternatively, it is interesting to consider the system with dynamical
quarks in the adjoint representation since adjoint quarks would
preserve the center symmetry. Actually the color structure of adjoint
quarks is identical to that of gluons. Thus the Polyakov loop serves
as an appropriate order parameter for the deconfinement transition
with respect to \textit{fundamental} quarks in a similar way as in the
pure gluonic theories. The chiral phase transition for dynamical
adjoint quarks and the deconfinement transition for static fundamental
quarks can be defined separately in this case. In the lattice
simulation with adjoint quarks (lattice aQCD), it has been found that
the critical temperature for the chiral restoration is distinctly
higher than that for the deconfinement transition \cite{kar99}. In
this paper we apply the Gocksch-Ogilvie model to investigate the
system with adjoint quarks as well.

This paper is organized as follows: We explain what the
Gocksch-Ogilvie model is with emphasis on its physical meaning in
Sec.~\ref{sec:model}. We give computational details in
App.~\ref{app:construction} additionally. In Sec.~\ref{sec:analytic}
we elaborate the mean field approximation employed in the present
analysis. Then in an analytic way we investigate the effects on the
chiral dynamics arising from the Polyakov loop behaviors.
Sec.~\ref{sec:numerical} is devoted to presenting the numerical
results and discussing their physical implications. Finally, we
conclude our model study in Sec.~\ref{sec:conclusions}.


\section{GOCKSCH-OGILVIE MODEL}
\label{sec:model}

The Gocksch-Ogilvie model is given by the effective action
\cite{goc85,fuk03}; (see App.~\ref{app:construction} for detail
derivations)
\begin{align}
 S_{\text{eff}}^{(r)}[L,\lambda] &=-J\sum_{\text{n.n.}}\mathrm{Tr_c}
  L(\vec{n})\mathrm{Tr_c}L^\dagger(\vec{m})+\frac{\text{dim}(r)}{2}
  \sum_{m,n}\lambda(n)V(n,m)\lambda(m) \notag\\
 &\qquad\qquad -\frac{N_{\text{f}}}{4}\sum_{\vec{n}}\mathrm{Tr_c}\ln
  \Bigl[\cosh(N_\tau E)+\frac{1}{2}\bigl(L^{(r)}+L^{(r)\dagger}\bigr)
  \Bigr],
\label{eq:GO_model}
\end{align}
where $L(\vec{n})$ is the Polyakov loop in the fundamental
representation defined on the lattice by
\begin{equation}
 L(\vec{n}) = \prod_{n_d=a}^{N_\tau a}U_d(\vec{n},n_d)
\end{equation}
in the $d$ dimensional space-time. $L^{(r)}$ denotes the Polyakov loop 
in the $r$ representation ($r=\text{fund.}$ for the fundamental
representation and $r=\text{adj.}$ for the adjoint one). The Polyakov
loop in the adjoint representation can be expressed in terms of the
fundamental ones as
\begin{equation}
 L_{ab}^{\text{(adj.)}}=2\mathrm{tr}\Bigl[t_a L t_b L^\dagger\Bigr],
\label{eq:adjoint}
\end{equation}
where the matrices $t_a$ form a fundamental representation of the
SU($N_{\text{c}}$) group and the indices $a,b$ run from $1$ to
$N_{\text{c}}^2-1$. The normalization is determined to satisfy
$\text{tr}[t_a t_b]=\delta_{ab}/2$.

In the effective action (\ref{eq:GO_model}), $\lambda(n)$ is the meson
field, $\mathrm{Tr_c}$ the trace with respect to the color indices,
and $N_{\text{f}}$ the number of flavors which is taken as
$N_{\text{f}}=2$ throughout this paper. $\text{dim}(r)$ is the number
of dimension in the $r$ representation of the SU($N_{\text{c}}$)
group, that is; $\text{dim}(\text{fund.})=N_{\text{c}}$ and
$\text{dim}(\text{adj.})=N_{\text{c}}^2-1$. The temperature $T$, the
meson hopping propagator $V(n,m)$, the strength of the nearest
neighbor interaction $J$, and the quasi-quark energy (constituent
quark mass) $E$ are defined respectively as
\begin{align}
 & T=\frac{1}{N_\tau a}, && V(n,m)=\frac{1}{2(d-1)}\sum_{\hat{j}}
  \bigl(\delta_{n,m+\hat{j}}+\delta_{n,m-\hat{j}}\bigr), \notag\\
 & J=\mathrm{e}^{-\sigma a/T}, && E=\sinh^{-1}\biggl(\sqrt{
  \frac{d-1}{2}}\lambda +m_q\biggr)
\label{eq:TVJE}
\end{align}
with $\hat{j}$ running from $1$ to $d-1$ only in the spatial
directions. $\sigma$ stands for the string tension fixed at the
empirical value at zero temperature, i.e.,
$\sigma=(425\,\text{MeV})^2$. The lattice spacing $a$ and the current
quark mass $m_q$ are treated as the model parameters.

\begin{figure}
\includegraphics[width=10cm]{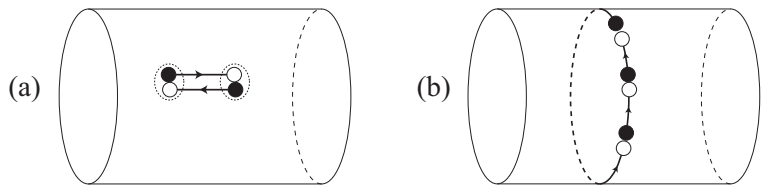}
\caption{Intuitive pictures for the effective action of the
Gocksch-Ogilvie model. The torus consists of the thermal $S^1$ and the
spatial $R^3$. The arrowed line represents the link variable. The
filled and blank circles are the quark and anti-quark excitations. (a)
In the spatial directions a quark and an anti-quark propagate as a
meson as a result of the group integration with respect to the spatial
link variables. (b) Thermal excitations of dynamical quarks are
allowed by the non-vanishing Polyakov loop at finite temperature.}
\label{fig:GO_model}
\end{figure}

Physical meaning for each term in the effective action
(\ref{eq:GO_model}) is straightforward to understand. The first term
is the gluonic contribution in terms of the Polyakov loop. An
order-disorder phase transition described by this gluonic action
signifies the transition from the center symmetric confined phase to
the symmetry broken deconfined phase. The next term involving
$\lambda(n)$ is the meson interaction term in the spatial direction
resulting from the group integration with respect to the spatial link
variables as shown in Fig.~\ref{fig:GO_model}~(a). The last
logarithmic term corresponds to the thermal excitation of dynamical
quarks as depicted intuitively in Fig.~\ref{fig:GO_model}~(b). This
interpretation will become tangible if once the Polyakov loop
$L^{(r)}$ is set to be unity by hand. The logarithmic term then takes
a familiar form of the free energy for fermionic ideal gases.

In the case of fundamental quarks the center symmetry is explicitly
broken by the thermal quark excitation with fractional baryon number.
Actually the expansion of the logarithmic quark contribution in the
power of the Polyakov loop is essentially equivalent to the hopping
parameter expansion \cite{gre84} and then the power of the Polyakov
loop counts the winding number of quark paths along the thermal torus,
namely, the quark number in the system. The leading contribution in
such an expansion in terms of the Polyakov loop is
$\sim\mathrm{Tr_c}L/\cosh(N_\tau E)$, which can be regarded as the
symmetry breaking term under an external magnetic-like field in
analogy to spin systems. The magnitude of the symmetry breaking
external field is specified by $1/\cosh(N_\tau E)$ hence it becomes
smaller as the constituent quark mass gets larger. Thus Satz's
argument on the relation between the Polyakov loop behavior and the
chiral dynamics is already embodied qualitatively in the
Gocksch-Ogilvie model in a natural way. Of course, the model will give
us deeper insight into the interplay between the deconfinement
transition and the chiral phase transition beyond qualitative
arguments, as we will see in the subsequent sections.

Now we shall remark about the applicable range in which the
Gocksch-Ogilvie model can make sense. The limit of the temperature
going to zero would render the model illegitimate because separate
treatments for the spatial directions and for the thermal direction as
illustrated in Fig.~\ref{fig:GO_model} can be allowed safely only at
high temperature. In order to take the limit of zero temperature
properly in the Gocksch-Ogilvie model, the group integration with
respect to the Polyakov loop, as performed in Ref.~\cite{dam86}, is
indispensable, which needs somewhat complicated algebraic procedures.
The interested temperature here is, as we will see later, around
$200\,\text{MeV}$ which corresponds to $N_\tau\sim2$ in the present
analysis. Thus we believe that it is well-founded to deal with the
thermal direction separately in the construction of the model. The
opposite limit of the temperature getting extremely high would make
the quantitative argument meaningless simply because the strong
coupling expansion does not work in a perturbative regime at such high
temperature. The Stefan-Boltzmann law, for example, cannot be derived
within the framework of the strong coupling expansion on the lattice.
In any case, as far as the region around the critical temperature is
concerned, we can presume that the Gocksch-Ogilvie model should be
applicable.


\section{ANALYTIC CALCULATIONS}
\label{sec:analytic}


\subsection{Mean-Field Approximation}

The dynamics of the Polyakov loop can be understood in analogy to spin
systems with the center symmetry. As usually performed in the
calculation in the spin system, we will adopt the mean field (Weiss)
approximation for the Polyakov loop. In this approximation, we take
account of the fluctuation of the individual Polyakov loop surrounded
by a constant mean field around it. The mean field action is taken
here as
\begin{equation}
 S_{\text{mf}}[L]=-\frac{x}{2}\sum_{\vec{n}}\bigl\{\mathrm{Tr_c}
  L(\vec{n})+\mathrm{Tr_c}L^\dagger(\vec{n})\bigr\},
\label{eq:mf_action}
\end{equation}
where $x$ is the mean field for the Polyakov loop and treated as a
variational parameter to be determined afterward from the extremal
condition on the free energy. As for the chiral order parameter,
$\lambda$ is simply regarded as a constant mean field denoted by
$\bar{\lambda}$. The mean field free energy is then given by
\cite{cre83}
\begin{align}
 \beta f_{\text{mf}}^{(r)}(x,\bar{\lambda}) &=-N^{-(d-1)}\Bigl(\langle
  -S_{\text{eff}}^{(r)}[L,\bar{\lambda}]+S_{\text{mf}}[L]
  \rangle_{\text{mf}}+\ln\int\mathcal{D}L\,
  \mathrm{e}^{-S_{\text{mf}}[L]}\Bigr) \notag\\
 &= \beta f_{\text{gluon}}(x)
  +\beta f_{\text{quark}}^{(r)}(x,\bar{\lambda}).
\end{align}
Here $\langle\cdots\rangle_{\text{mf}}$ means the expectation value
with the weight specified by the mean field action. $N^{(d-1)}$ is the
spatial volume. Roughly speaking, the first term
$\langle S_{\text{eff}}^{(r)}\rangle_{\text{mf}}$ corresponds to the
internal energy tending to make the system ordered and remaining terms
are the entropy tending to make the system disordered. 

The gluonic part of the free energy can be written as
\begin{equation}
 \beta f_{\text{gluon}}(x) =-2(d-1)J\biggl(\frac{\mathrm{d}}
  {\mathrm{d}x}\ln I(x)\biggr)^2+x^2\frac{\mathrm{d}}{\mathrm{d}x}
  \biggl(\frac{1}{x}\ln I(x)\biggr),
\label{eq:free_gluon}
\end{equation}
where $I(x)$ is defined by \cite{kog82}
\begin{equation}
 I(x) =\int\mathrm{d}L\,\exp\biggl[\frac{x}{2}\mathrm{Tr_c}\bigl(
  L+L^\dagger\bigr)\biggr] = \sum_m \det I_{m-i+j}(x)
\label{eq:I}
\end{equation}
with $I_n(x)$ representing the modified Bessel function of the first
kind. The quark part of the free energy in the $r$ representation is
written for $N_{\text{f}}=2$ as
\begin{equation}
 \beta f_{\text{quark}}^{(r)}(x,\bar{\lambda}) =\frac{\text{dim}(r)
  N_\tau}{2}\bar{\lambda}^2-\frac{\text{dim}(r)}{2}\ln\bigl[
  \cosh(N_\tau E)\bigr]-\frac{1}{2}\frac{\tilde{I}^{(r)}(x;
  \cosh(N_\tau E))}{I(x)}.
\label{eq:free_quark}
\end{equation}
$\tilde{I}^{(r)}(x;\alpha)$ is given in the fundamental and in the
adjoint representation respectively by
\begin{align}
 \tilde{I}^{\text{(fund.)}}(x;\alpha) &=\frac{1}{N_{\text{c}}!}
  \sum_{a=1}^{N_{\text{c}}}\sum_{m=-\infty}^\infty
  \epsilon_{i_1\cdots i_{N_{\text{c}}}}
  \epsilon_{j_1\cdots j_{N_{\text{c}}}}I_{m-i_1+j_1}(x)\cdots
  \tilde{I}_{m-i_a+j_a}(x;\alpha)\cdots
  I_{m-i_{N_{\text{c}}}+j_{N_{\text{c}}}}(x), \notag\\
 \tilde{I}_n(x;\alpha) &=\int_0^{2\pi}\frac{\mathrm{d}\phi}{2\pi}
  \ln\Bigl[1+\frac{\cos\phi}{\alpha}\Bigr]\cos(n\phi)\,
  \mathrm{e}^{x\cos\phi},
\label{eq:til_I_1}
\end{align}
and
\begin{align}
 \tilde{I}^{\text{(adj.)}}(x;\alpha) &=\frac{1}{N_{\text{c}}!}
  \sum_{a\neq b=1}^{N_{\text{c}}}\sum_{m=-\infty}^\infty
  \epsilon_{i_1\cdots i_{N_{\text{c}}}}
  \epsilon_{j_1\cdots j_{N_{\text{c}}}}I_{m-i_1+j_1}(x)\cdots
  \tilde{I}_{m-i_b+j_b}^{m-i_a+j_a}(x;\alpha)\cdots
  I_{m-i_{N_{\text{c}}}+j_{N_{\text{c}}}}(x) \notag\\
 &\qquad +(N_{\text{c}}-1)\ln\Bigl[1+\frac{1}{\alpha}\Bigr]I(x),
  \notag\\
 \tilde{I}_{n}^{m}(x;\alpha) &=\int_0^{2\pi}\frac{\mathrm{d}\phi_a
  \mathrm{d}\phi_b}{(2\pi)^2}\ln\Bigl[1+\frac{\cos(\phi_a-\phi_b)}
  {\alpha}\Bigr]\cos(m\phi_a+n\phi_b)\,
  \mathrm{e}^{x(\cos\phi_a+\cos\phi_b)} \notag\\
 &=\int_0^{2\pi}\frac{\mathrm{d}\phi}{2\pi}\ln\Bigl[1+\frac{\cos\phi}
  {\alpha}\Bigr]\cos\Bigl(\frac{m-n}{2}\phi\Bigr)I_{m+n}\Bigl(
  2x\cos\frac{\phi}{2}\Bigr).
\label{eq:til_I_2}
\end{align}
In Eqs.~(\ref{eq:til_I_1}) and (\ref{eq:til_I_2}) the integration in
terms of $\phi$ originates from the thermal excitation of fundamental
quarks. Since the excitation of an adjoint quark consists of a
fundamental quark and anti-quark pair (see Eq.~(\ref{eq:adjoint})),
two integration variables $\phi_a$ and $\phi_b$ appear in
Eq.~(\ref{eq:til_I_2}).

Now that we have the actual expressions for the mean field free energy
given by Eqs.~(\ref{eq:free_gluon}) and (\ref{eq:free_quark}), all we
have to do is find the minimum of the free energy with respect to the
variational parameters $x$ and $\bar{\lambda}$. Before addressing the
numerical analyses, we shall look into the analytic expressions of the
free energy further in the next subsection.


\subsection{Chiral Symmetry Breaking in the Confined Phase}

It would be worth considering the particular cases where we can
manipulate the mean field free energy in analytic ways. In this
and the next subsections we take the chiral limit, namely, $m_q=0$ and
discuss what becomes of the chiral symmetry in some particular cases.

An interesting limit is that the mean field for the Polyakov loop is
forced to be zero by hand, i.e., $x=0$. This limit readily corresponds
to the confined phase for adjoint quarks because the Polyakov loop
serves as the order parameter for the deconfinement transition in the
system with adjoint quarks preserving the center symmetry. To be
interesting, even for fundamental quarks, the limit of $x=0$ should
indicate the confined phase. The reason is as follows; the group
integration with respect to the Polyakov loop would pick out only
color singlet combinations in the integrand. In the limit of $x=0$,
i.e., without any external source from the mean field, the expansion
of the logarithmic contribution in terms of the Polyakov loop gives
rise to only center symmetric combinations. Actually, the condition of
$x=0$ leads to the description in the canonical ensemble with the
quark triality fixed at zero (see Ref.~\cite{fuk03_2} and references
therein). The criterion for the deconfinement transition, therefore,
regains its meaning based on the center symmetry and it follows that
the vanishing Polyakov loop signifies the confined phase even for
fundamental quarks.

For the quark part of the free energy, taking $x=0$ makes the
expressions of Eqs.~(\ref{eq:til_I_1}) and (\ref{eq:til_I_2})
simplified as
\begin{align}
 \tilde{I}^{\text{(fund.)}}(0;\alpha) &=N_{\text{c}}
  \tilde{I}_0(0;\alpha)-2(-1)^{N_{\text{c}}}
  \tilde{I}_{N_{\text{c}}}(0;\alpha) \\
 \tilde{I}^{\text{(adj.)}}(0;\alpha) &=N_{\text{c}}(N_{\text{c}}-1)
  \tilde{I}_0(0;\alpha)-\sum_{a\neq b=1}^{N_{\text{c}}}
  \tilde{I}_{a-b}(0;\alpha).
\end{align}
Then we can make use of the following integrations \footnote{The
author thanks T.~Kunihiro for the derivation of these integration
formulae.},
\begin{align}
 \tilde{I}_0(0;\alpha) &=\int_0^{2\pi}\frac{\mathrm{d}\phi}{2\pi}\ln
  \Bigl[1+\frac{\cos\phi}{\alpha}\Bigr]=\ln\biggl[\frac{\alpha
  +\sqrt{\alpha^2-1}}{2\alpha}\biggr] \notag\\
 \tilde{I}_n(0;\alpha) &=\int_0^{2\pi}\frac{\mathrm{d}\phi}{2\pi}\ln
  \Bigl[1+\frac{\cos\phi}{\alpha}\Bigr]\cos(n\phi)=-\frac{1}{n}\bigl(
  \sqrt{\alpha^2-1}-\alpha)^n \qquad(n\neq0)
\end{align}
to come by the expanded form of the free energy in terms of
$\bar{\lambda}$ as \footnote{The expansion given by Eq.~(16) of
Ref.~\cite{fuk03} should be corrected;
$N_{\text{c}}$ replaced by $N_{\text{c}}-2$ and a factor $1/\sqrt{2}$
inserted. Thus the linear term disappears in the case of
$N_{\text{c}}=2$.}
\begin{align}
 \beta f_{\text{mf}}^{\text{(fund.)}}(0,\bar{\lambda}) &\simeq
  \text{(const.)}-\sqrt{\frac{d-1}{2}}\biggl(\frac{N_{\text{c}}}{2}-1
  \biggr)N_\tau |\bar{\lambda}|+\mathrm{O}(\bar{\lambda}^2),
\label{eq:conf_chiral1} \\
 \beta f_{\text{mf}}^{\text{(adj.)}}(0,\bar{\lambda}) &\simeq
  \text{(const.)}-\sqrt{\frac{d-1}{2}}\biggl[\frac{N_{\text{c}}^2}{2}
  \biggr] N_\tau|\bar{\lambda}| +\mathrm{O}(\bar{\lambda}^2),
\label{eq:conf_chiral2}
\end{align}
where $[\cdots]$ in Eq.~(\ref{eq:conf_chiral2}) denotes the greatest
integer function which gives the greatest integer less than or equal
to the number. To derive the above expressions, we have substituted
$\alpha=\cosh(N_\tau E)\simeq 1+(d-1)N_\tau^2\bar{\lambda}^2/4$.

Here we should emphasize the appearance of a linear term of
$|\lambda|$ which stems from the expansion of $\sqrt{\alpha^2-1}$.
Such a linear term always causes instability around the chiral
symmetric vacuum ($\bar{\lambda}=0$). In other words, we have shown
from Eqs.~(\ref{eq:conf_chiral1}) and (\ref{eq:conf_chiral2}) that the
Gocksch-Ogilvie model with the mean field approximation has the
property that \textit{the chiral symmetry is inevitably broken
spontaneously in the confined phase} for both fundamental and adjoint
quarks. This property is in accordance with the general theoretical
requirement.


\subsection{Chiral Restoration Temperature}

We can now consider the opposite situation where the Polyakov loop is
almost trivial in the deconfined phase and the chiral restoration
occurs continuously. Then, as discussed in Refs.~\cite{goc85,fuk03},
we can acquire the chiral restoration temperature by the quadratic
term of the free energy expanded in terms of the chiral order
parameter $\bar{\lambda}$. Supposing that the mean field for the
Polyakov loop takes a substantially finite value, let us assume that
the Polyakov loop can be approximated simply by
$L_{ij}\simeq v\cdot\delta_{ij}$ and
$L^{\text{(adj.)}}_{ab}\simeq v^2\cdot\delta_{ab}$, though this
approximation breaks the unitarity of the matrices $L$ and
$L^{\text{(adj.)}}$. The free energies for fundamental and adjoint
quarks are expanded up to the quadratic order as
\begin{align}
 \beta f_{\text{quark}}^{\text{(fund.)}}(L=v,\bar{\lambda}) &\simeq
  \text{(const.)}+\frac{N_{\text{c}}N_\tau}{2}\biggl[1-
  \frac{(d-1)N_\tau}{4(v+1)}\biggr]\bar{\lambda}^2
  +\mathrm{O}(\bar{\lambda}^4), \\
 \beta f_{\text{quark}}^{\text{(adj.)}}(L=v,\bar{\lambda}) &\simeq
  \text{(const.)}+\frac{(N_{\text{c}}^2-1)N_\tau}{2}\biggl[1-
  \frac{(d-1)N_\tau}{4(v^2+1)}\biggr]\bar{\lambda}^2
  +\mathrm{O}(\bar{\lambda}^4).
\end{align}
From these expansions, we can immediately have the chiral restoration
temperatures;
\begin{equation}
 T_\chi^{\text{(fund.)}} =\frac{d-1}{4(v+1)}a^{-1}, \qquad
 T_\chi^{\text{(adj.)}} =\frac{d-1}{4(v^2+1)}a^{-1}.
\label{eq:chiral_Tc}
\end{equation}
In contrast to the lattice observation \cite{kar99}, the chiral
restoration temperature for adjoint quarks is not changed radically
from that for fundamental quarks. From the analytic expressions for
the critical temperature we can see the tendency that the chiral
restoration should be hindered as the system approaches the confined
phase where $v\sim\langle\mathrm{Tr_c}L\rangle$ gets smaller, though
the above expressions are only valid for $v\sim1$.


\subsection{Modification to the Deconfinement Temperature}

Although the angle integrations appearing in Eqs.~(\ref{eq:til_I_1})
and (\ref{eq:til_I_2}) seem too complicated to accomplish
analytically, the expansion in terms of $1/\alpha$, which will be
acceptable when the quark mass is large, enables us to write down
$\tilde{I}_n(x;\alpha)$ and $\tilde{I}_n^m(x;\alpha)$ immediately (see
also App.~\ref{app:expansion}). Such an expansion in the leading order
of $1/\alpha$ and a further expansion in terms of $x$ as well yield
the following expressions of the free energy for $N_{\text{c}}=3$;
\begin{align}
 \beta f_{\text{gluon}}(x) &\simeq -2(d-1)J\Bigl(\frac{x}{2}
  +\frac{x^2}{8}-\frac{5x^4}{384}-\frac{x^5}{256}\Bigr)^2
  +\frac{x^2}{4}+\frac{x^3}{12}-\frac{x^5}{96}-\frac{5x^6}{1536}, \\
 \beta f_{\text{quark}}^{\text{(fund.)}}(x,\bar{\lambda}) &\simeq
  \frac{3}{2}N_\tau\bar{\lambda}^2-\frac{3}{2}\ln\cosh(N_\tau E)
  -\frac{1}{2\cosh(N_\tau E)}\Bigl(\frac{x}{2}+\frac{x^2}{8}
  -\frac{5x^4}{384}-\frac{x^5}{256}\Bigr),
\label{eq:ex_fund} \\
 \beta f_{\text{quark}}^{\text{(adj.)}}(x,\bar{\lambda}) &\simeq
  4N_\tau\bar{\lambda}^2-4\ln\cosh(N_\tau E)
  -\frac{1}{2\cosh(N_\tau E)}\Bigl(\frac{x^2}{4}+\frac{x^3}{12}
  -\frac{x^5}{96}\Bigr).
\label{eq:ex_adj}
\end{align}

\begin{figure}
\includegraphics[width=3.5cm]{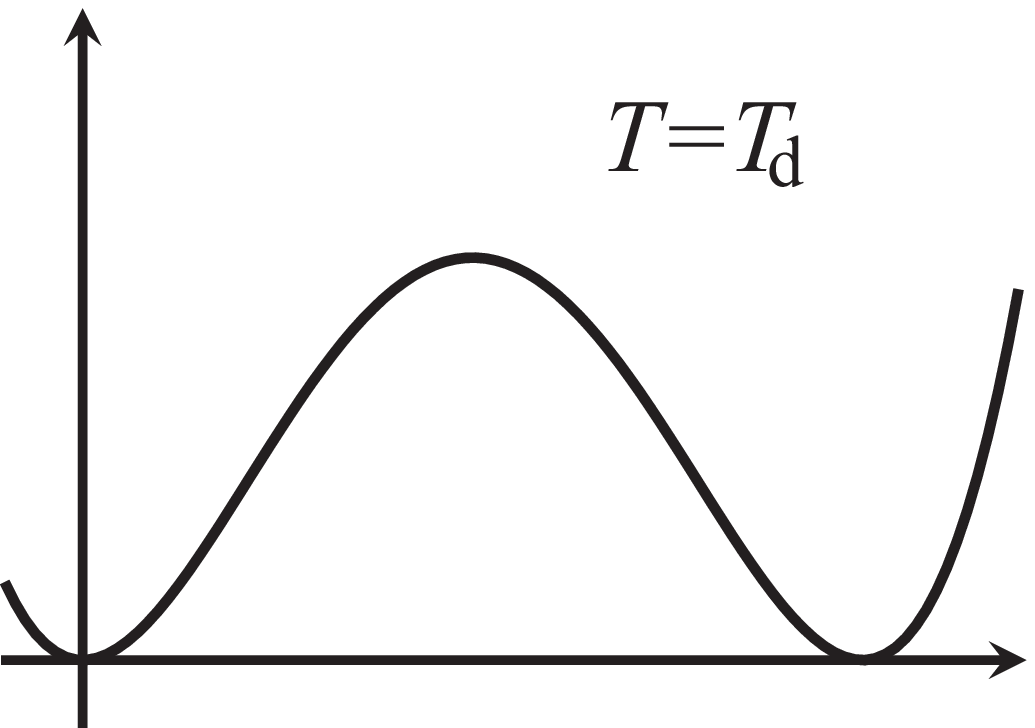} \hspace{8mm}
\includegraphics[width=3.5cm]{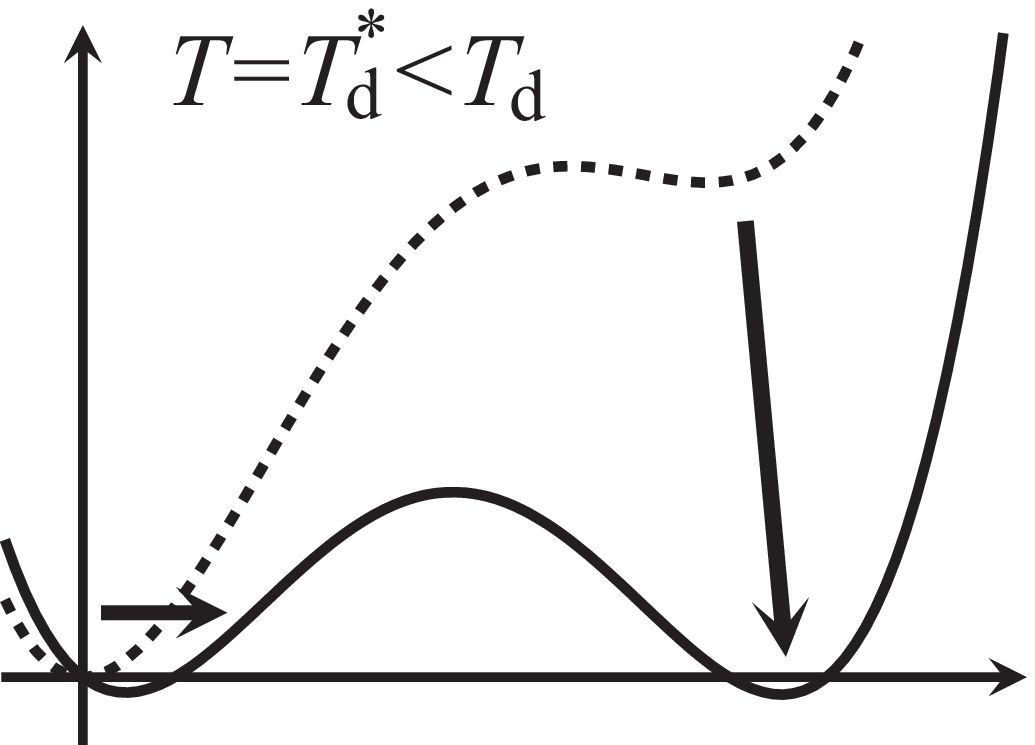} \hspace{8mm}
\includegraphics[width=3.5cm]{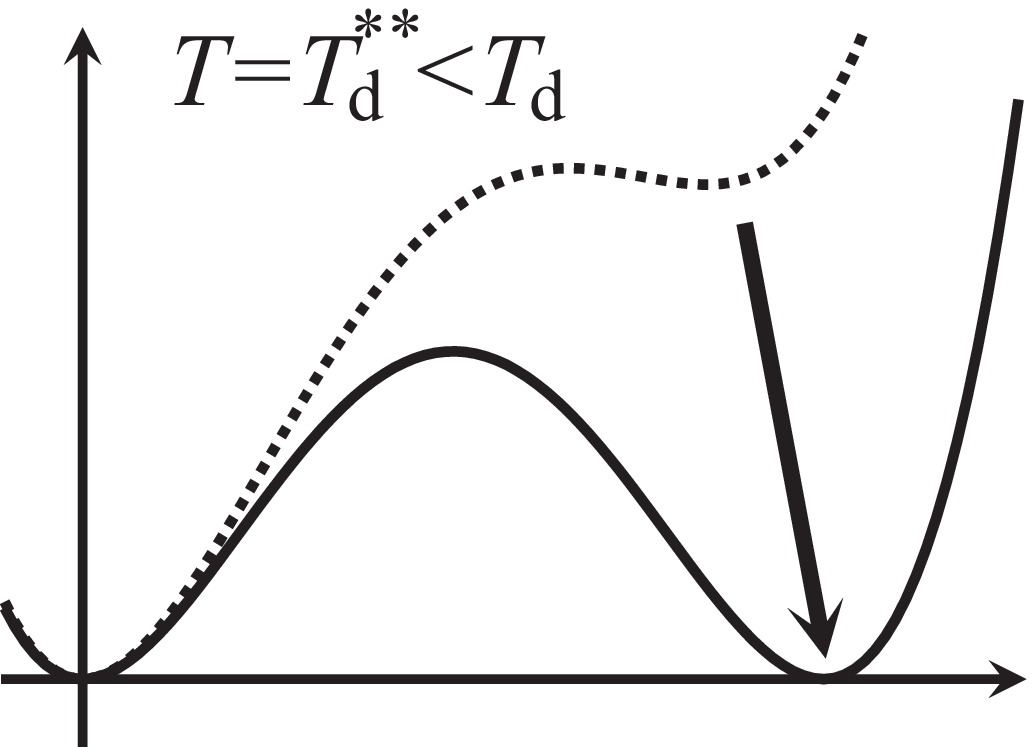}
\caption{Schematic figures to show the modification to the free
energy. The left figure is for the first order transition at
$T=T_{\text{d}}$ in the pure gluonic case. The transition temperature
is lowered from $T_{\text{d}}$ to $T_{\text{d}}^\ast$ and the minimum
around the origin is shifted by the linear term for fundamental quarks
(middle figure). The transition temperature is also lowered to
$T_{\text{d}}^{\ast\ast}$ by the quadratic term for adjoint quarks
(right figure).}
\label{fig:poten}
\end{figure}

The gluonic part of the free energy always has a minimum around $x=0$
because the expansion starts from the order of $x^2$. Thus it can
describe the first order deconfinement transition from $x=0$ to some
finite $x$ at the deconfinement temperature $T_{\text{d}}$ as shown
schematically in the left figure of Fig.~\ref{fig:poten}. The absence
of the linear term of $x$ reflects the center symmetric property.

The expansion (\ref{eq:ex_fund}) for fundamental quarks clearly shows
that the quark contribution breaks the center symmetry through the
linear term of $x$. In the presence of such a linear term, the minimum
around $x=0$ in the gluonic free energy is shifted to a finite value
of $x$. As far as only the linear term, which is negative now, is
concerned, the free energy will be more decreased at larger $x$. As a
result, the first order deconfinement transition temperature is
lowered from $T_{\text{d}}$ to $T_{\text{d}}^\ast<T_{\text{d}}$ by the
effect of the explicit breaking of the center symmetry, if the first
order transition persists. The intuitive picture is depicted in the
middle figure of Fig.~\ref{fig:poten}. As a matter of fact, the first
order transition occurs even in the presence of dynamical quarks with
$m_q\gtrsim300\,\text{MeV}$ for Parameter~I and
$m_q\gtrsim700\,\text{MeV}$ for Parameter~II (the meaning of the model
parameters is given in the next section).

In Eq.~(\ref{eq:ex_adj}), on the other hand, the leading term from the
logarithmic contribution is quadratic in $x$ and the minimum around
$x=0$ is not altered then. This is because of the center symmetry
remaining for adjoint quarks. When we focus attention on the leading
term, as in the above case for fundamental quarks, the negative
contribution of order $x^2$ in the quark free energy will reduce the
free energy more for larger $x$. Therefore, as shown in the right
figure of Fig.~\ref{fig:poten}, the first
order deconfinement transition temperature is also lowered to
$T_{\text{d}}^{\ast\ast}<T_{\text{d}}$ in the presence of adjoint
quarks.

The strength of how much the deconfinement temperature is lowered,
i.e., $T_{\text{d}}-T_{\text{d}}^\ast$ or
$T_{\text{d}}-T_{\text{d}}^{\ast\ast}$, is specified by
$\sim1/\cosh(N_\tau E)$. As long as the current quark mass, $m_q$, is
smaller than the scalar condensate $\bar{\lambda}$, or as long as the
constituent quark mass $Ea^{-1}$ is smaller than the temperature
$T=1/N_\tau a$, the modification to the deconfinement temperature will
be almost insensitive to $m_q$. This anticipation, in fact, has been
observed in the lattice aQCD simulations \cite{kar99} and will be
confirmed in our numerical results in the next section.


\section{NUMERICAL RESULTS AND DISCUSSIONS}
\label{sec:numerical}

The parameters inherent in the Gocksch-Ogilvie model are the lattice
spacing (cut-off) $a$ and the current quark mass $m_q$. In our
previous work \cite{fuk03}, $a$ and $m_q$ are chosen to fit the masses
of pions and $\rho$ mesons. If $m_q$ is much smaller than $a^{-1}$
then $a$ is essentially determined by the mass difference between
pions and $\rho$ mesons and $m_q$ is fixed by the condition to
reproduce the pion mass. The numerical values of the model parameters
in this case are listed as Parameter~I in Table~\ref{tab:parameter}.
The deconfinement transition in a pure gluonic system occurs at
$2(d-1)J_{\text{c}}=0.806$, that is; $T_{\text{d}}=208\,\text{MeV}$
with these values of the model parameters. This value of
$T_{\text{d}}$ seems to be rather low as compared with the empirical
value ($T_{\text{d}}=270\,\text{MeV}$) found in the lattice
simulations. Since the transition temperatures are crucial rather than
the hadron spectrum here, we will try another parameter set,
Parameter~II, in this paper; instead of using the $\rho$ meson mass,
we can fix the lattice spacing $a$ so as to reproduce the empirical
value of the deconfinement transition temperature by substituting
$J_{\text{c}}$ and the empirical $T_{\text{d}}$ for
Eq.~(\ref{eq:TVJE}). We summarize our choice of the model parameters
and the resultant outputs in Table~\ref{tab:parameter}. As we will
discuss later, these two parameter sets (Parameter~I and Parameter~II)
happen to correspond to two typical cases for the realization of the
simultaneous phase transitions.

\begin{table}
\begin{tabular}{c|ccccccc}
\hline\hline
 & $a^{-1}$ & $m_q$ & $m_\pi$ & $m_\rho$ &
  $\Psi$ & $T_{\text{d}}$ & $T_{\text{c}}$ \\ \hline
Parameter~I & $432\,\text{MeV}$ & $5.7\,\text{MeV}$ &
  $^\ast 140\,\text{MeV}$ & $^\ast 770\,\text{MeV}$ &
  $-(342\,\text{MeV})^3$ & $208\,\text{MeV}$ & $\sim187\,\text{MeV}$
  \\
Parameter~II & $333\,\text{MeV}$ & $7.4\,\text{MeV}$ &
  $^\ast 140\,\text{MeV}$ & $597\,\text{MeV}$ &
  $-(263\,\text{MeV})^3$ & $^\ast 270\,\text{MeV}$ &
  $\sim230\,\text{MeV}$ \\
\hline\hline
\end{tabular}
\caption{Model parameters employed in the numerical calculations and
the resultant outputs. $\ast$-quantities are used as inputs in order
to fix $a$ and $m_q$. $\Psi$ is the chiral condensate.}
\label{tab:parameter}
\end{table}

It should be noted that we have determined the model parameters by
using the expressions of the hadron masses at zero temperature
according to Ref.~\cite{klu83}, instead of those achieved in the zero
temperature limit of the Gocksch-Ogilvie model. We briefly review the
necessary expressions at zero temperature with modifications for
$N_{\text{f}}$ flavors in App.~\ref{app:zero}. In order to take the
zero temperature limit of the Gocksch-Ogilvie model we have to
integrate out the Polyakov loop variable properly. It is, however, a
somewhat subtle procedure and thus we choose to use the would-be
solutions at zero temperature from the beginning. Actually the
numerical value of the model parameters fixed in this way turns out to
be almost the same as that fixed by the approximate zero temperature
limit (without the Polyakov loop integration) of the Gocksch-Ogilvie
model as actually adopted in Ref.~\cite{goc85}.

\begin{figure}
\includegraphics[width=7cm]{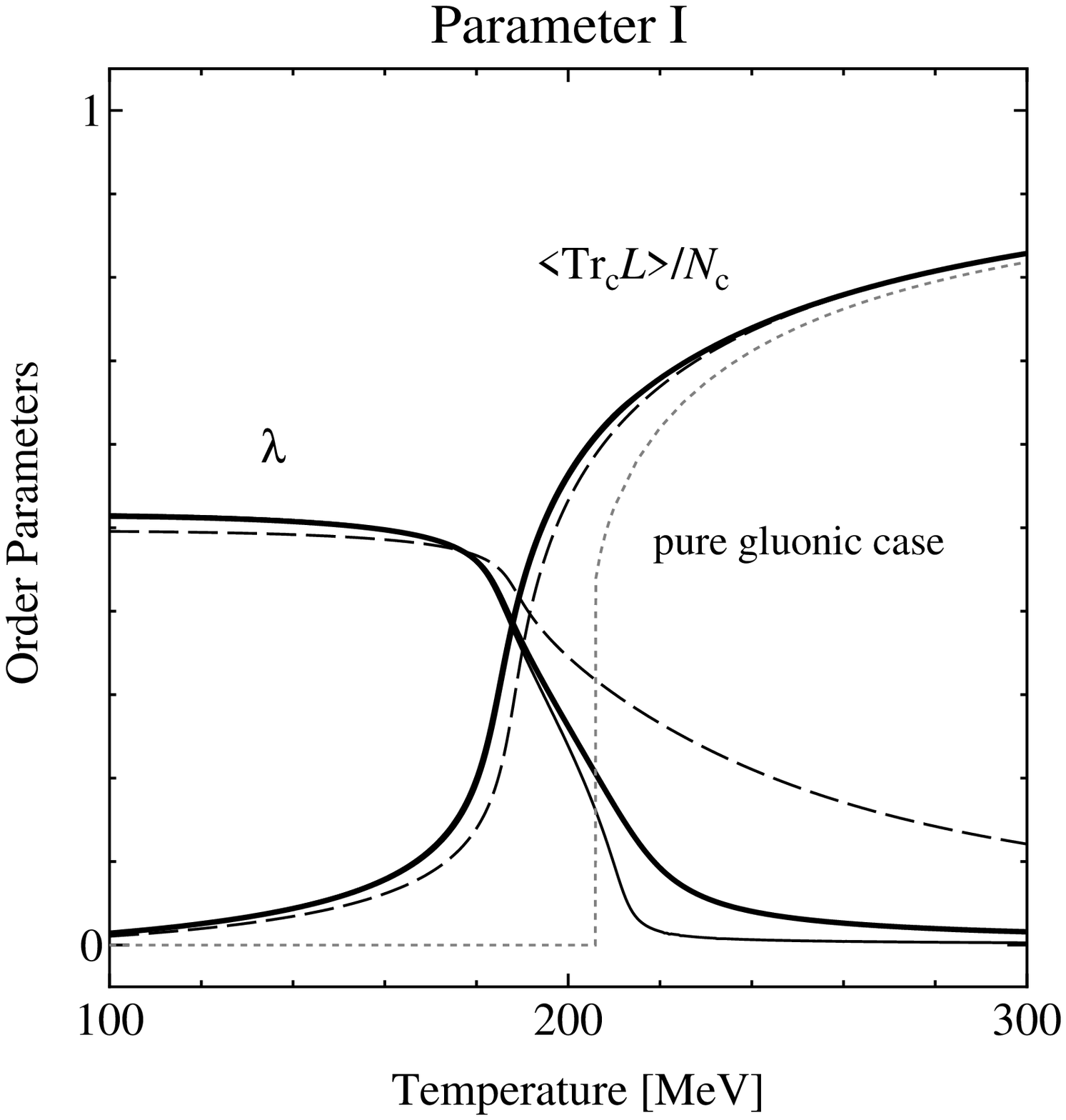} \hspace{5mm}
\includegraphics[width=7cm]{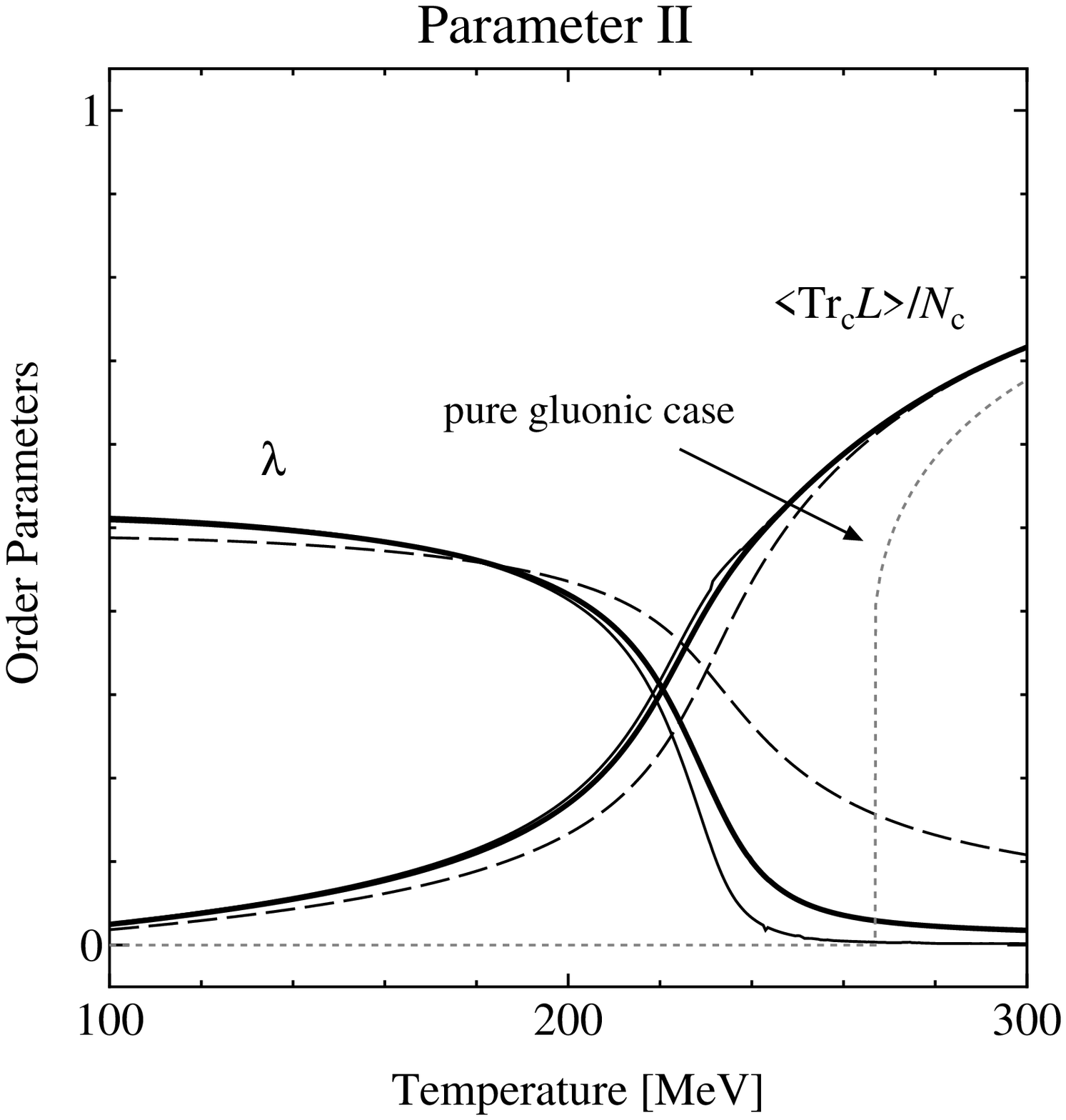}
\caption{Behaviors of the order parameters as functions of the
temperature for $N_{\text{c}}=3$. The thick solid curve is for
$m_q=5.7\,\text{MeV}$ in the left figure (Parameter~I) and for
$m_q=7.4\,\text{MeV}$ in the right figure (Parameter~II). The thin
solid curve is for $m_q=1\,\text{MeV}$ and the dashed curve for
$m_q=50\,\text{MeV}$.}
\label{fig:fund}
\end{figure}

The numerical results for $N_{\text{c}}=3$ are presented in
Fig.~\ref{fig:fund} together with the results for several different
choices of the current quark mass. In the presence of quark
contributions, the expectation value of the Polyakov loop is not equal
to $x$ itself, though in the pure gluonic case the extremal condition
on the free energy ensures that the mean field for the Polyakov loop
is simply provided by $x$. From Eqs.~(\ref{eq:mf_action}) and
(\ref{eq:I}) we can have the following relation;
\begin{equation}
 \frac{1}{2N_{\text{c}}}\langle\mathrm{Tr_c}L+\mathrm{Tr_c}L^\dagger
  \rangle =\frac{1}{N_{\text{c}}}\frac{\mathrm{d}I(x)}{\mathrm{d}x},
\end{equation}
from which we have calculated the expectation value of the Polyakov
loop. As for the chiral order parameter, we simply use the scalar
condensate $\bar{\lambda}$, for we can show that it is proportional
to the conventional order parameter, i.e., the chiral condensate
$\Psi=\langle\bar{\chi}\chi\rangle$. Further details are given in
App.~\ref{app:chiral}.

\begin{figure}
\includegraphics[width=7cm]{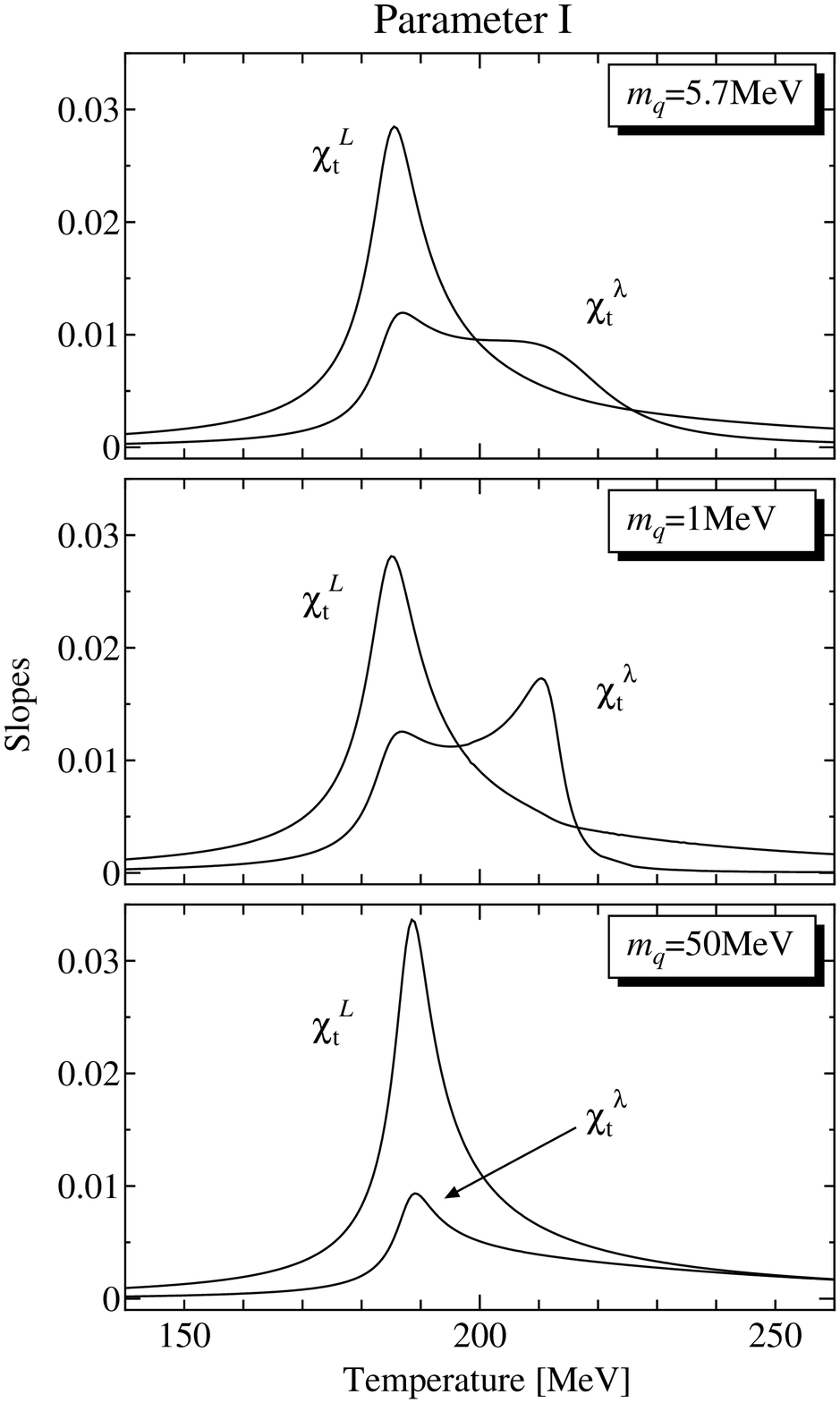} \hspace{5mm}
\includegraphics[width=7cm]{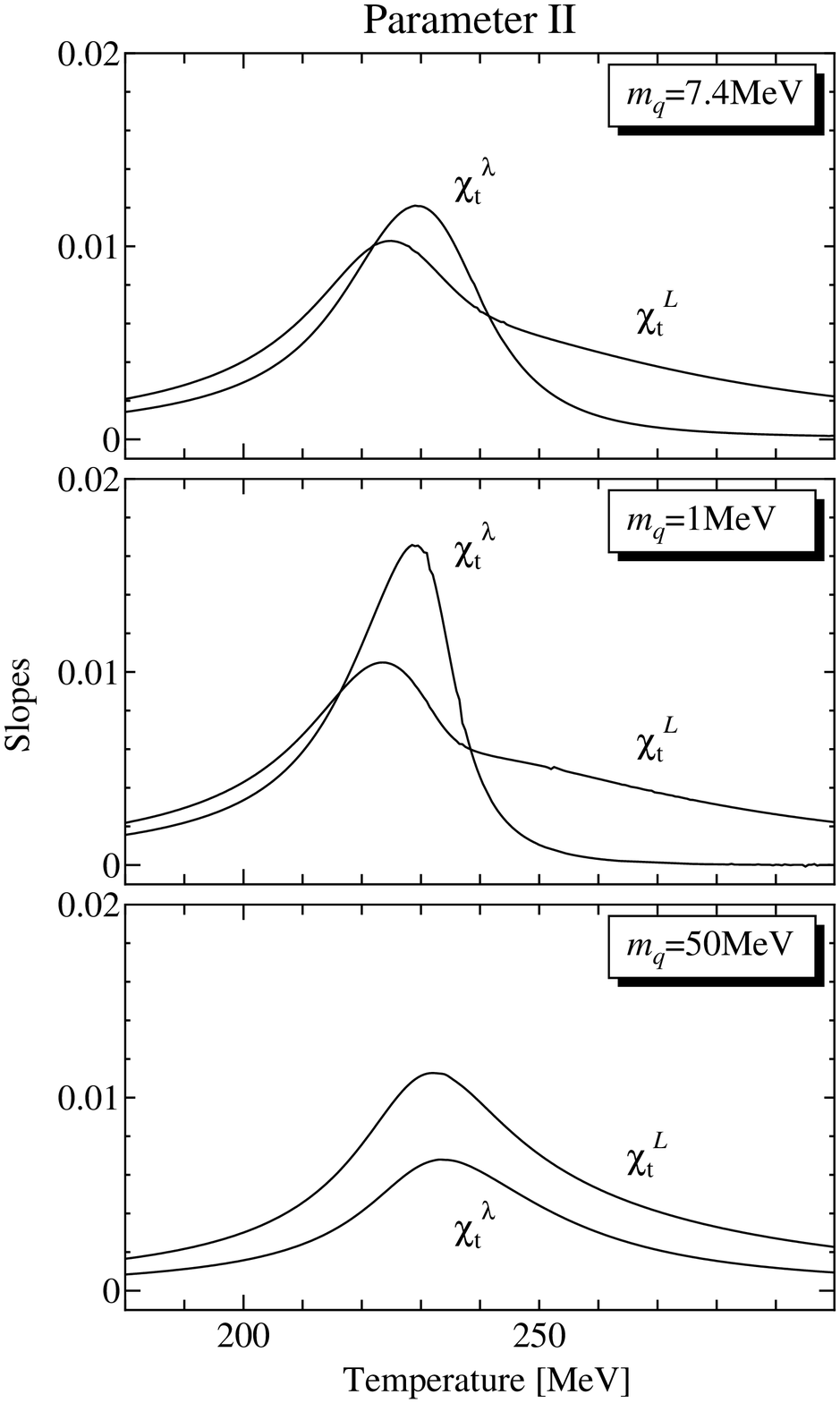}
\caption{Slopes (temperature derivatives) of the order parameters.}
\label{fig:slope}
\end{figure}

In both cases of Parameter~I and Parameter~II, we can see the order
parameters for deconfinement and chiral restoration indicating
crossover behaviors almost simultaneously. Although the numerical
results in Fig.~\ref{fig:fund} might seem just supportive for Satz's
argument, the relation between the Polyakov loop and the chiral order
parameter has richer contents in itself, which can be deduced from
Fig.~\ref{fig:slope}.

In Fig.~\ref{fig:slope} we plot the temperature derivative of the
order parameters defined by
\begin{equation}
 \chi_{\text{t}}^L =\frac{1}{N_{\text{c}}}\frac{\partial
  \langle\mathrm{Tr_c}L\rangle}{\partial T}, \qquad
 \chi_{\text{t}}^\lambda =-\frac{\partial\bar{\lambda}}{\partial T},
\end{equation}
which are nothing but the slope of the curves drawn in
Fig.~\ref{fig:fund}.

In both cases of the parameter sets, the pseudo-critical temperature,
$T_{\text{c}}$, inferred from the peak of $\chi_{\text{t}}^L$ is
almost identical to that from the peak of $\chi_{\text{t}}^\lambda$.
Surely the coincidence of the pseudo-critical temperatures found in
the lattice QCD simulation can be reproduced in the slope of the order
parameters within the present model study. It is interesting to see
that the physical implication for the case of Parameter~I is different
from that for the case of Parameter~II.

For $m_q=1\,\text{MeV}$ with Parameter~I, two peaks emerge apparently
in the behavior of $\chi_{\text{t}}^\lambda$. In this case one peak
around $T_{\text{c}}\simeq 187\,\text{MeV}$ corresponds to the remnant
of the first order deconfinement transition rather than the chiral
phase transition as conjectured formerly in Ref.~\cite{fuk03}. The
other around $T\simeq210\,\text{MeV}$ has the origin from the second
order chiral phase transition, which can be also confirmed by the thin
curve behavior in Fig.~\ref{fig:fund}. Therefore the coincidence of
the pseudo-critical temperatures signifies not the chiral phase
transition but the induced instability of the chiral order triggered
by the deconfinement transition.

For $m_q=1\,\text{MeV}$ with Parameter~II, on the other hand, there
appears only one peak around $T_{\text{c}}\simeq230\,\text{MeV}$.
The peak height of $\chi_{\text{t}}^\lambda$ around $T_{\text{c}}$
becomes larger as the current quark mass is lowered, which is peculiar
to the genuine phase transition. Therefore, this peak is considered as
resulting from the second order chiral phase transition and the
transition temperature, in fact, agrees with that read from the thin
curve behavior in Fig.~\ref{fig:fund}. Thus the coincidence of the
pseudo-critical temperatures in this case can be understood according
to Satz's argument.

Which transition governs the simultaneous jump around the
pseudo-critical temperature relies on the magnitude relation between
the chiral restoration temperature $T_\chi$ and the lowered
deconfinement temperature $T_{\text{d}}^\ast$. In the case of
Parameter~I, $T_{\text{d}}^\ast<T_\chi$ is realized to lead to, so to
speak, the \textit{deconfinement dominance} around $T_{\text{c}}$. In
reality, because two peaks are not observed in the lattice QCD
simulation, a more relevant situation is to be described by
Parameter~II where $T_{\text{d}}^\ast>T_\chi$ or
$T_{\text{d}}^\ast\simeq T_\chi$ would hold. The former
$T_{\text{d}}^\ast>T_\chi$ means the \textit{chiral dominance} around
$T_{\text{c}}$ and then the critical phenomena are to be characterized
by the critical exponents associated with the chiral phase
transition. However, because of the general property that the chiral
phase transition must take place at higher temperature than the
deconfinement transition does, as we have explained in the previous
section, the latter $T_{\text{d}}^\ast\simeq T_\chi$ is likely to be
realized. If it is the case, the chiral phase transition might be so
affected by the remnant of the first order deconfinement transition
that the critical phenomena peculiar to the second order transition
would be obscured. This can be one possible account for the
unsatisfactory agreement between the anticipation based on the
universality argument and the critical exponents measured in the
lattice QCD simulations \cite{kar94}.

\begin{figure}
\includegraphics[width=7.5cm]{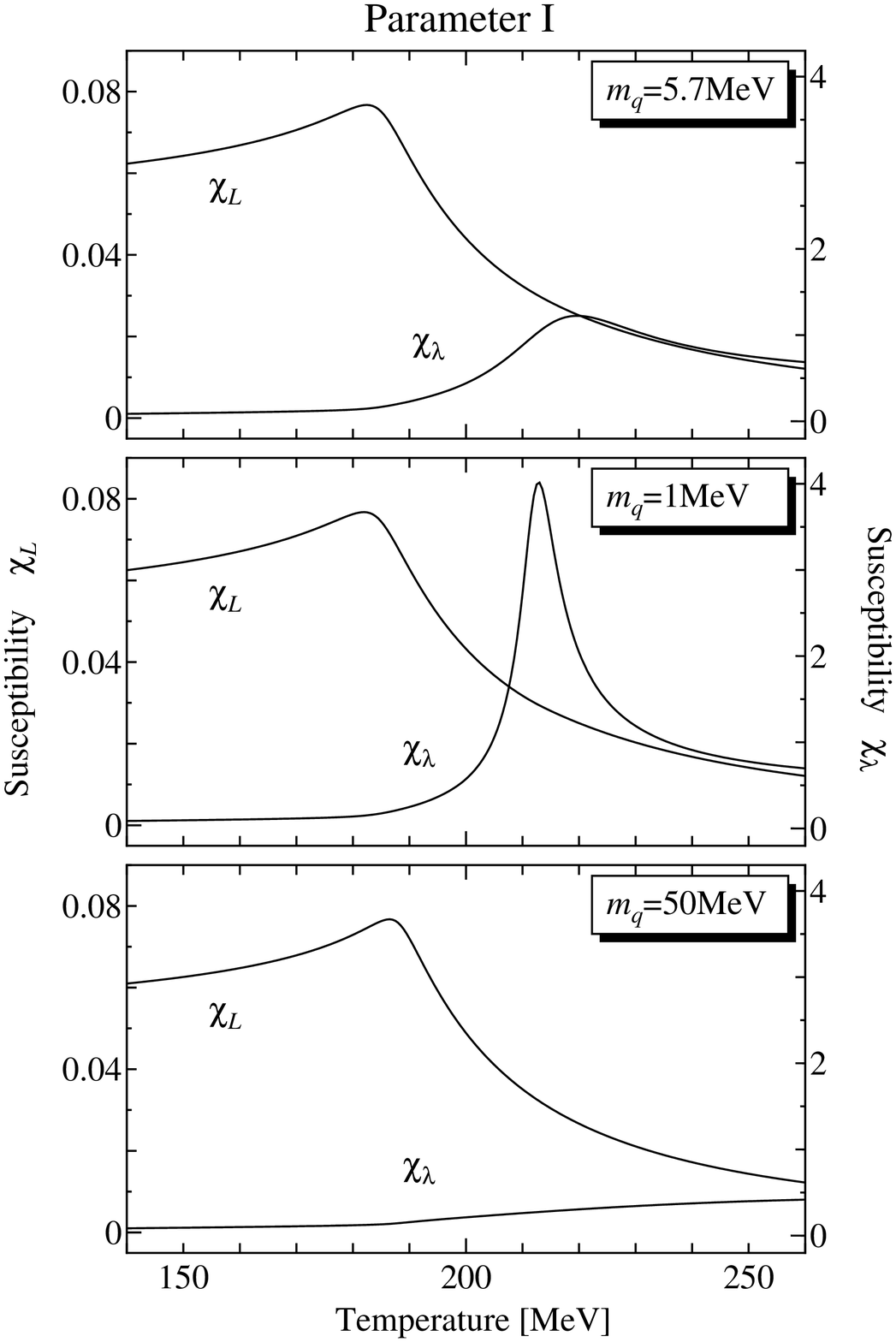} \hspace{5mm}
\includegraphics[width=7.5cm]{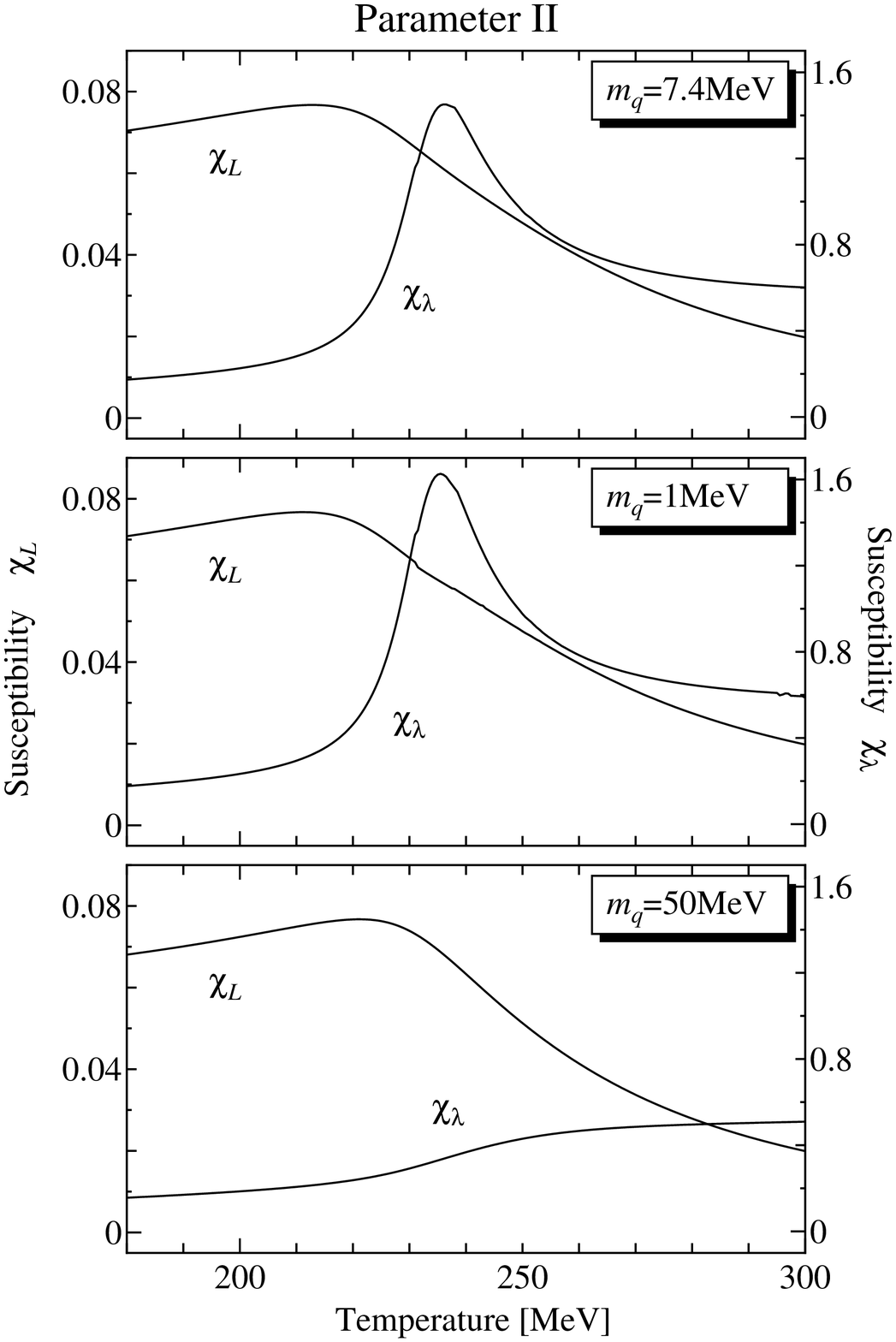}
\caption{Susceptibilities with respect to the order parameters.}
\label{fig:suscept}
\end{figure}

Fig.~\ref{fig:suscept} is the figure for the susceptibility with
respect to the order parameters. The susceptibilities are here to be
calculated by
\begin{align}
 \chi_L &=\frac{1}{N_{\text{c}}^2}\Bigl(\langle(\mathrm{Tr_c}L)^2
  \rangle-\langle\mathrm{Tr_c}L\rangle^2\Bigr)
 =\frac{1}{N_{\text{c}}^2}\frac{\mathrm{d}^2}{\mathrm{d}x^2}
  \ln I(x),
\label{eq:sus_pol} \\
 \chi_\lambda &=\langle\lambda^2\rangle-\langle\lambda\rangle^2
 =\biggl(\frac{\partial^2 \beta f_{\text{mf}}^{\text{(fund.)}}}
  {\partial\bar{\lambda}^2}\biggr)^{-1}.
\label{eq:sus_lam}
\end{align}
The definition of $\chi_\lambda$ seems to be different from the
conventional chiral susceptibility $\chi_\Psi=\langle(\bar{\chi}\chi
)^2\rangle-\langle\bar{\chi}\chi\rangle^2$, but $\chi_\lambda$ is
almost proportional to $\chi_\Psi$ near the critical temperature as
explained in App.~\ref{app:chiral}.

The genuine susceptibilities appear to be substantially different from
the slopes and do not indicate any clear coincidence of the location
of peaks. This is partly because of the very difference between the
slope and the susceptibility and partly because of the approximation.

The slope $\chi_{\text{t}}^\lambda$ and the susceptibility
$\chi_\lambda$ are certainly sensitive to the second order chiral
phase transition, and yet they can have different behaviors. In the
chiral limit for a clear example, the slope vanishes above the
critical temperature $T_\chi$, while the susceptibility becomes finite
or even larger than the magnitude below $T_\chi$. Actually simple
mean field arguments typically give twice larger amplitude for the
susceptibility above $T_\chi$ than that below $T_\chi$.

The slope and the susceptibility relevant to the Polyakov loop, on the
other hand, do not have to enhance due to the first order
deconfinement transition. Therefore, for Parameter~I, we have no
reason to expect any coincidence from the beginning because of the
deconfinement dominance where the Polyakov loop behavior has little
information on the second order transition. The Polyakov loop behavior
is still smooth even for Parameter~II. We consider that it is because
the simple ansatz of the mean field action (\ref{eq:mf_action}) would
miss the significant part of the coupling to the symmetry breaking
contribution through which the coincidence is brought about. This
ansatz can incorporate, at most, a linear term of an external
magnetic-like field and thus the non-linearity leading to the fact
that the chiral phase transition temperature must be higher than the
deconfinement transition temperature would be lost as a result of the
approximation. Hence we cannot see the coincidence of the peak
location even for Parameter~II.

Now we would comment upon one interesting possibility: The results
from the lattice QCD simulation suggest that the situation of
Parameter~II is more realistic, and nevertheless, it would be
appealing to consider some modification or artificial tuning in the
lattice simulation, by which the case of Parameter~I is realized. Then
we can make use of the lattice simulation directly to investigate the
nature of the deconfinement transition around $T_{\text{c}}$ and the
nature of the chiral phase transition around $T_\chi>T_{\text{c}}$
separately.

\begin{figure}
\includegraphics[width=7cm]{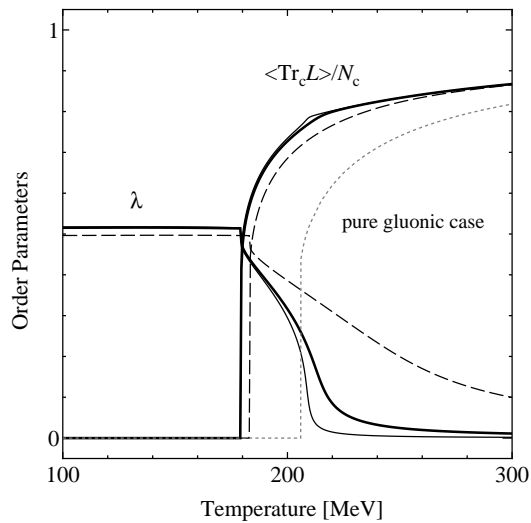}
\caption{Behaviors of the order parameters for adjoint quarks with
Parameter~I. The thick solid curve is for $m_q=5.7\,\text{MeV}$ and
the thin solid and the dashed curves for $m_q=1\,\text{MeV}$ and for
$m_q=50\,\text{MeV}$ respectively.}
\label{fig:adjoint}
\end{figure}

Finally we show the numerical results for adjoint quarks with
Parameter~I in Fig.~\ref{fig:adjoint}. The calculation has been done
with the function $\tilde{I}_n^m(x;\alpha)$ expanded in terms of
$1/\alpha$ up to third order, for the numerical integration is not
impossible but somewhat tough. We give the explicit expressions and
verify the reliability of such an expansion in
App.~\ref{app:expansion}. As expected, the deconfinement transition of
first order persists clearly and the chiral phase transition occurs
distinctly at higher temperature than the deconfinement transition
does.

The global features consist well with the lattice aQCD results. The
deconfinement temperature is lowered from
$T_{\text{d}}=208\,\text{MeV}$ to
$T_{\text{d}}^{\ast\ast}\simeq180\,\text{MeV}$. For various choices of
$m_q$, the lowered deconfinement temperature $T_{\text{d}}^{\ast\ast}$
is almost the identical, as we have already discussed in the previous
section.

The behavior of the chiral order parameter, as pointed out in
Ref.~\cite{kar99}, changes radically at $T=T_{\text{d}}$; the current
quark mass dependence is not manifested until the temperature exceeds
the deconfinement temperature. This feature can be understood
qualitatively from Eq.~(\ref{eq:chiral_Tc}). While the Polyakov loop
is zero, the chiral order parameter behaves as a function of the
temperature as if the chiral phase transition would take place at much
higher temperature. After the deconfinement transition, the chiral
restoration comes to affect the chiral order parameter behavior and,
as a result, the quark mass dependence becomes apparent then.


\section{CONCLUSIONS}
\label{sec:conclusions}

We investigated the relation between the deconfinement transition and
the chiral phase transition at finite temperature in the
Gocksch-Ogilvie model with fundamental or adjoint quarks incorporated.
We shall enumerate our findings here;

\begin{enumerate}
\item
We have shown in an analytic way that, in the confined phase with the
vanishing Polyakov loop, the chiral symmetry should be spontaneously
broken at any temperature in the Gocksch-Ogilvie model for both
fundamental and adjoint quarks.
\item
We discussed qualitatively how the quark contribution to the effective
action can lower the deconfinement transition temperature.
\item
In the case of fundamental quarks we have acquired the behavior of the
order parameters as functions of the temperature and their slopes by
differentiating with respect to the temperature. Then we found the
peaks of the slope for the Polyakov loop and for the chiral order
parameter located at the same pseudo-critical temperature.
\item
Physical implication of the coincidence of the pseudo-critical
temperatures has proved to be twofold: One is the deconfinement
dominance where the remnant of the first oder deconfinement transition
would govern the behavior of the order parameters around
$T_{\text{c}}$. The other is the chiral dominance where the second
order chiral phase transition would be responsible for the realization
of the simultaneous jump of the order parameters around
$T_{\text{c}}$. We speculate that the real world corresponds to an
intermediate admixture between the deconfinement dominance and the
chiral dominance.
\item
In the case of adjoint quarks we have found that the deconfinement
transition is seen clearly of first order and that the chiral phase
transition of second order occurs at higher temperature than the
deconfinement transition does. It is apparent that the deconfinement
temperature is substantially lowered from $T_{\text{d}}$ to
$T_{\text{d}}^{\ast\ast}$, as expected, and the difference
$T_{\text{d}}-T_{\text{d}}^{\ast\ast}$ is insensitive to the current
quark mass $m_q$ as far as $m_q$ is smaller than the scalar condensate
(i.e.\ the constituent quark mass) or the temperature.
\end{enumerate}

Some comments are in order. We will address our conjecture here that
the finding 1 would be generalized so that we can prove the chiral
symmetry breaking in the confined phase only using the property of the
group integration, or, the Haar measure. Also it would be interesting
to make sure whether 1 is realized in the lattice simulation. Such
simulations will be feasible, in principle, if once we can examine the
chiral symmetry breaking with a constraint imposed on the expectation
value of the Polyakov loop, i.e., $\langle\mathrm{Tr_c}L\rangle=0$.

As for 3, we must improve the mean field approximation, especially the
ansatz of the mean field action, to reproduce the coincidence of the
pseudo-critical temperatures in the genuine susceptibility. This is a
task for the future.

It would be fascinating from 4 to consider the possibility of the
deconfinement dominance realizing in the lattice simulation because we
can investigate the nature of the deconfinement around
$T_{\text{c}}\simeq T_{\text{d}}^\ast$ and the nature of the chiral
restoration around $T_\chi>T_{\text{c}}$ separately. In order to
achieve such a situation, we have to make some modifications to the
lattice QCD either to lower $T_{\text{d}}^\ast$ or to raise $T_\chi$.
The problem of how to put the idea into practice is beyond the scope
of the present paper.

Although it is rather academic, we finally remark about the possible
relevance to physics on the critical end point (CEP). Recently the
CEP, that is the terminal point of the first order transition, is paid
much attention to in the realistic QCD phase diagram \cite{ste98}. The
location of the CEP is determined in part in the lattice QCD
simulation at finite temperature and density \cite{fod02}. The CEP can
also emerge as the strange quark mass is varied in the three-flavor
chiral phase transition \cite{gav94}.

Here we will draw attention to another chance to be confronted with
the CEP, namely, the point where the deconfinement transition ceases
to be first order with massive dynamical quarks. Our numerical
analysis in the Gocksch-Ogilvie model predicts that the critical quark
mass $m_q^{\text{c}}\sim300\,\text{MeV}$ for Parameter~I and
$m_q^{\text{c}}\sim700\,\text{MeV}$ for Parameter~II. Actually there
exist some lattice measurements to determine the critical mass of the
CEP: In Ref.~\cite{has83} $m_q^{\text{c}}\simeq 900\,\text{MeV}$ is
concluded, while Ref.~\cite{att88} reads
$m_q^{\text{c}}\simeq 400\,\text{MeV}$ in more improved analyses with
two-flavor quarks. The results in Ref.~\cite{att88} is consistent with
our estimate of $m_q^{\text{c}}$. In the one-flavor QCD simulation
$m_q^{\text{c}}\simeq 1.4\,\text{GeV}$ is reported \cite{ale99}. As
for the three-flavor case \cite{kar01}, the situation becomes somewhat
subtle because not only the deconfinement transition with heavy quarks
but also the chiral phase transition with light quarks can be of first
order. We anticipate that the property of the CEP in the deconfinement
transition might be intriguing from the point of view of the glueball
physics at finite temperature \cite{ish02,hat03}.

In view of the prosperous descriptions as listed above, we can regard
the Gocksch-Ogilvie model as a promising implement to afford a
conceptual clue to the relation between the deconfinement transition
and the chiral phase transition. It would be intriguing to make
progress in applying the model to the system at finite density.

\begin{acknowledgments}
The author, who is supported by the Japan Society for the Promotion of
Science for Young Scientists, would like to thank T.~Hatsuda and
S.~Sasaki and Y.~Nishida for discussions. He thanks T.~Kunihiro for
his sincere encouragement and warm hospitality at Yukawa Institute for
Theoretical Physics.
\end{acknowledgments}

\appendix


\section{CONSTRUCTION OF THE GOCKSCH-OGILVIE MODEL}
\label{app:construction}

In order to make this paper self-contained, we shall explain how to
construct the Gocksch-Ogilvie model explicitly by following the
derivation of Refs.~\cite{klu83,goc85}. This would be useful to
clarify what approximation is implicitly presumed. The model is
formulated on the lattice. The strong coupling expansion is readily
feasible on the lattice to lead to the quark confining property, i.e.,
the area law of the Wilson loop. Although the real world is actually
in the weak coupling regime, numerical studies in the lattice QCD have
revealed that even the strong coupling expansion should be likely to
give a plausible picture relevant to the real world. The spontaneous
breaking of the chiral symmetry can be described on the lattice as
well by means of the double expansion of strong coupling and large
dimensionality \cite{klu83,dam86}. Thus we can expect that an
effective action suitable for discussing the relation between the
Polyakov loop and the chiral order parameter will be available if once
we bring together those assets on the lattice. This is the starting
point to construct the Gocksch-Ogilvie model.

We adopt the staggered fermion in order to approach the chiral limit
in a plain manner and also in order to take advantage of the
simplicity in regard to the spin structure. The action on the lattice
is given by the gluonic part;
\begin{equation}
 S_{\text{G}}[U]=\frac{2N_{\mathrm{c}}}{g^2}\sum_{n,(\mu,\nu)}
  \biggl\{1-\frac{1}{N_{\mathrm{c}}}\text{Re}\mathrm{Tr_c}
  U_{\mu\nu}(n)\biggr\},\quad
  U_{\mu\nu}(n)=U_\nu^\dagger(n)U_\mu^\dagger(n+\hat{\nu})
   U_\nu(n+\hat{\mu})U_\mu(n)
\end{equation}
for $N_{\mathrm{c}}\ge3$ and the staggered quark part;
\begin{align}
 S_{\text{F}}[U_d,U_i,\bar{\chi},\chi] &=m_q
  \sum_n\bar{\chi}(n)\chi(n)+  \frac{1}{2}\sum_{n,\mu>0}
  \eta_\mu(n)\bar{\chi}(n)\bigl\{U_\mu(n)\chi(n+\hat{\mu})
  -U_\mu^\dagger(n-\hat{\mu})\chi(n-\hat{\mu})\bigr\} \notag\\
 &=m_q \sum_n\bar{\chi}(n)\chi(n)+  \frac{1}{2}\sum_n
  \eta_d(n)\bar{\chi}(n)\bigl\{U_d(n)\chi(n+\hat{d})
  -U_d^\dagger(n-\hat{d})\chi(n-\hat{d})\bigr\} \notag\\
 &\qquad\qquad -\sum_{n,i\neq d}\mathrm{Tr_c}\Bigl\{
  \bar{\Phi}_i(n)U_i(n)+\Phi_i(n)U_i^\dagger(n)\Bigr\},
\end{align}
where
\begin{align}
 \bigl\{\Phi_i(n)\bigr\}_{ab} &=\frac{1}{2}\eta_i(n+\hat{i})
  \bar{\chi}_b(n+\hat{i})\chi_a(n), \notag\\
 \bigl\{\bar{\Phi}_i(n)\bigr\}_{ab} &=-\frac{1}{2}\eta_i(n)
  \bar{\chi}_b(n)\chi_a(n+\hat{i}).
\end{align}
We have divided the quark action into one part with $U_i$ and the
other part without $U_i$ for later convenience. $a$, $b$ run in the
color space. We can handle the quark action in the adjoint
representation by replacing the link variables in the covariant
derivatives by
\begin{equation}
 U_{ab}^{\text{(adj.)}}=2\mathrm{tr}\Bigl[t_a U t_b U^\dagger
  \Bigr].
\end{equation}
Roughly speaking, the adjoint quark is like a color octet fermionic
meson composed of a quark and an anti-quark in the fundamental
representation ($\mathbf{8}$ out of
$\mathbf{3}\otimes\bar{\mathbf{3}}=\mathbf{1}\oplus\mathbf{8}$). We
will write the link variable as $U^{(r)}$ in the $r$ representation.

In order to attain the effective action in terms of the Polyakov loop
and the quark field, we must integrate over the spatial link variable
$U_i$. Then the effective partition function is written as
\begin{align}
 Z_{\text{eff}}^{(r)}[U_d,\chi,\bar{\chi}]
 &=\int\mathcal{D}U_i\, \mathrm{e}^{-S_{\text{G}}[U_d,U_i]
  -S_{\mathrm{F}}[U,\chi,\bar{\chi}]} \notag\\
 &=\exp\biggl[ m_q \sum_n\bar{\chi}(n)\chi(n)+ \frac{1}{2}\sum_n
  \eta_d(n)\bar{\chi}(n)\bigl\{U_d^{(r)}(n)\chi(n+\hat{d})
  -U_d^{(r)\dagger}(n-\hat{d})\chi(n-\hat{d})\bigr\}\biggr] \notag\\
 &\qquad\qquad \times\int\mathcal{D}U_i\,
  \exp\biggl[ -S_{\text{G}}[U_d,U_i]+\sum_{n,i\neq d}
  \mathrm{Tr_c}\Bigl\{\bar{\Phi}_i(n)U_i^{(r)}(n)+\Phi_i(n)
  U_i^{(r)\dagger}(n)\Bigr\}\biggr].
\label{eq:eff_part}
\end{align}

In the strong coupling expansion, the spatial plaquette can be
neglected, which can be perceived more on the asymmetric lattice
\cite{bil96}. An extreme case of such asymmetric lattice methods is
the Hamiltonian formalism where only the temporal plaquette remains in
the strong coupling limit. Then the $U_i$ integration part can be
expanded as
\begin{equation}
 \int\mathcal{D}U_i\,\sum_{p,q}\frac{1}{p!q!}\Biggl[\frac{1}{g^2}
  \sum_{n,i\neq d}\mathrm{Tr_c}\Bigl\{U_{id}(n)+U_{id}^\dagger(n)
  \Bigr\}\Biggr]^p \Biggl[\sum_{n,i\neq d}\mathrm{Tr_c}\Bigl\{
  \bar{\Phi}_i U_i^{(r)}(n)+\Phi_i(n)U_i^{(r)\dagger}(n)\Bigr\}\Biggr]^q.
\label{eq:pq_expand}
\end{equation}
The Taylor expansion of the gluonic part (series in $p$) obviously
corresponds to the strong coupling ($1/g^2$) expansion. As we explain
soon below, the Taylor expansion of the quark part (series in $q$)
corresponds to the large dimensional ($1/d$) expansion. We specify the
order of the double expansion as follows; first, we perform the large
dimensional expansion. Then, in each expanded term we take the leading
contribution of the strong coupling expansion.

\begin{figure}
\includegraphics[width=4cm]{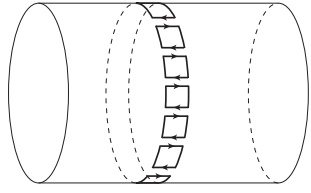}
\caption{Graphical representation of the configuration contributing to
the effective action in the lowest order of the strong coupling. The
torus stands for the thermal (temporal) $S^1$ and the spatial $R^3$.}
\label{fig:effect}
\end{figure}

For the zeroth order ($q=0$) term of the $1/d$ expansion, the leading
contribution left after the $U_i$ integration derives from the
configuration with the plaquettes wrapping around the thermal
(temporal) torus ($p=N_\tau$), as depicted in Fig.~\ref{fig:effect}.
As a result, the $U_i$ integration for the $q=0$ term is \cite{pol82}
\begin{equation}
 1+\biggl(\frac{1}{g^2 N_{\mathrm{c}}}\biggr)^{N_{\mathrm{c}}}
  \sum_{\text{n.n.}}\mathrm{Tr_c}L(\vec{n})
  \mathrm{Tr_c}L^\dagger(\vec{m}).
\label{eq:eff_action_g}
\end{equation}

The next contribution comes from the second order ($q=2$) term. Since
two link variables in the covariant derivative can form a color
singlet pair, the leading contribution in the strong coupling
expansion is simply the zeroth ($p=0$) term. As a result, we have
\begin{align}
 &\int\mathcal{D}U_i\,\Biggl[\sum_{n,i\neq d}\mathrm{Tr_c}\Bigl\{
  \bar{\Phi}_i U_i^{(r)}(n)+\Phi_i(n)U_i^{(r)\dagger}(n)\Bigr\}
  \Biggr]^2 \notag\\
 =&\sum_{n,i\neq d}\frac{1}{\text{dim}(r)}\mathrm{Tr_c}
  \bigl\{\bar{\Phi}_i(n)\Phi_i(n)\bigr\}
 =\sum_{n,i\neq d}\frac{\text{dim}(r)}{2(d-1)}M(n)M(n+\hat{i}),
\label{eq:dim_exp}
\end{align}
where the mesonic composite field is defined by
\begin{equation}
 M(n)=\frac{1}{\text{dim}(r)}\sqrt{\frac{d-1}{2}}
  \sum_{a=1}^{\text{dim}(r)}\chi_a(n)\bar{\chi}_a(n).
\end{equation}
In Eq.~(\ref{eq:dim_exp}) the dimension number of the $r$
representation is denoted by $\text{dim}(r)$ as in the text. It should
be noted that, for adjoint quarks, only such a mesonic interaction as
$\mathrm{Tr_c}\{\bar{\Phi}_i \Phi_i\}$ survives not only because of
the group integration with respect to $U_i$ but also because of the
properties of the Grassmann variables $\chi$ and $\bar{\chi}$.

At this stage it is easy to understand why such an expansion in $q$
provides the large dimensional expansion. Since the meson should
propagate with finite amplitude in the large dimensional limit, we
must properly normalize $M(n)$ so as to have the factor $1/(d-1)$ in
front of the term $M(n)M(n+\hat{i})$ ($i$ runs from $1$ to $d-1$).
As a result, each time $\Phi_i$, or $M$, appears in a term, the order
of the term in the $1/d$ expansion should be lowered. Thus the Taylor
expansion of the exponential of the quark action amounts to the
expansion in $1/d$.

We combine Eqs.~(\ref{eq:eff_part})$-$(\ref{eq:dim_exp}) and
exponentiate the expanded terms. Then, using the identity,
\begin{equation}
 \exp\biggl[\text{dim}(r)\sum_{m,n}\frac{1}{2}M(n)V(n,m)M(m)\biggr]
 =\int\mathcal{D}\lambda\,\exp\Biggl[-\text{dim}(r)\sum_{m,n}
  \biggl\{\frac{1}{2}\lambda(n)V(n,m)\lambda(m)-\lambda(n)V(n,m)M(m)
  \biggr\}\Biggr],
\label{eq:identity}
\end{equation}
we can reach the effective partition function;
\begin{align}
 Z_{\text{eff}}^{(r)}[U_d,\lambda] &=\int\mathcal{D}\chi
  \mathcal{D}\bar{\chi}\,\exp\Biggl[-S_{\text{eff}}^{\text{(G)}}[L]-
  \frac{\text{dim}(r)}{2}\sum_{m,n}\lambda(n)V(n,m)\lambda(m) \notag\\
 &\qquad -\frac{1}{2}\sum_{n}\eta_d(n)\Bigl\{\bar{\chi}(n)
  U_d^{(r)}(n)\chi(n+\hat{d})-\bar{\chi}(n+\hat{d})
  U_d^{(r)\dagger}(n)\chi(n)\Bigr\} \notag\\
 &\qquad\quad -\sum_n \bar{\chi}(n_d)\Biggl\{\sqrt{\frac{d-1}{2}}
  \sum_m\lambda(m)V(m,n)+m_q \Biggr\}\chi(n_d)\Biggr].
\label{eq:integ_chi}
\end{align}
From Eq.~(\ref{eq:eff_action_g}), the effective action in terms of the
Polyakov loop, $S_{\text{eff}}^{\text{(G)}}[L]$, is given by
\begin{equation}
 S_{\text{eff}}^{\text{(G)}}[L]=-J\sum_{\text{n.n.}}
  \mathrm{Tr_c}L(\vec{n})\mathrm{Tr_c}L^\dagger(\vec{m}).
\end{equation}
where $J$ is given by
\begin{equation}
 J=\biggl(\frac{1}{g^2 N_{\mathrm{c}}}\biggr)^{N_\tau}
  =\mathrm{e}^{-\sigma a/T}.
\label{eq:temperature}
\end{equation}
Here we have used the familiar expression for the string tension in
the strong coupling expansion, i.e.,
$\sigma a^2=\ln[g^2 N_{\mathrm{c}}]$. We will fix the lattice spacing
$a$ as a cut-off parameter and vary the temperature by changing the
number of lattice sites in the thermal direction, $N_\tau$.

In Eq.~(\ref{eq:integ_chi}) the hopping propagator of the meson fields
$V(n,m)$ is defined in Eq.~(\ref{eq:TVJE}). Since the quark part takes
a bilinear form in Eq.~(\ref{eq:integ_chi}), we can perform the
integration with respect to the quark fields explicitly under the
approximation that $\lambda$ is \textit{independent} of $n_d$ (time).
This approximation is indeed reasonable as long as we employ the mean
field approximation and treat $\lambda$ as a constant, in the same
sense as the sigma meson field in the relativistic nuclear many-body
system. We would emphasize, however, that this approximation might be
invalid when the meson fluctuation is concerned. In other words, we
should not use the action of the Gocksch-Ogilvie model to estimate the
meson fluctuation (loop) effect.

Then after the integration we can reach the Gocksch-Ogilvie model
given in Eq.~(\ref{eq:GO_model}) by using
$S_{\text{eff}}^{(r)}=-\ln Z_{\text{eff}}^{(r)}$. We should remark
that $N_{\text{f}}$ in the effective action (\ref{eq:GO_model}) is the
number of flavors introduced from the phenomenological point of
view. We cannot avoid having four degenerate flavors per a single
component of the staggered fermion. The logarithmic contribution in
Eq.~(\ref{eq:GO_model}) results from the quark integration in
Eq.~(\ref{eq:integ_chi}). The coefficient $N_{\text{f}}/4$ in front of
the logarithmic term effectively adjusts the number of flavors in the
integration with respect to the quark fields $\chi$.

It is worth mentioning that $\sigma$ in Eq.~(\ref{eq:temperature}) is
constant and fixed at the empirical value at zero temperature. The
string tension of the linear potential defined by the Polyakov loop
correlation function, nevertheless, depends on the temperature and
decreases with the temperature increasing. In reality the string
tension should have temperature dependence in addition due to the
vibrations of the flux tube. The higher temperature the system is
heated up at, the more furious the vibration becomes. Such effects of
the string vibration are to be described by the higher order
contributions in the strong coupling expansion. If we take account of
the spatial link variables, or the magnetic plaquettes in the
Hamiltonian formalism, to exceed the leading order contribution of
Eq.~(\ref{eq:eff_action_g}), we can incorporate temperature dependence
coming from the string vibration in principle. Since the essence on
the deconfinement transition is already concentrated in the effective
action~(\ref{eq:eff_action_g}), we simply neglect any spatial
fluctuation in the present model.

\begin{figure}
\includegraphics[width=3.5cm]{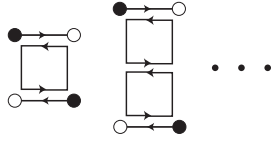}
\caption{Examples of neglected excitations as a result of the
approximation. The vertical and horizontal lines represent the
temporal (thermal) and spatial link variables respectively.}
\label{fig:fat}
\end{figure}

If we proceeded to the higher order contribution of the strong
coupling expansion, we would have the fat meson contributions as
shown in Fig.~\ref{fig:fat}. Neglecting such fat excitations implies
that we neglect the structural change of mesons in the temporal
direction. In other words, the meson hopping in the spatial direction
have no extent in the temporal direction and is always
color-confined.

\section{CHIRAL CONDENSATE}
\label{app:chiral}

The chiral condensate can be calculated as
\begin{equation}
 \Psi =\langle\bar{\chi}\chi\rangle=-\frac{1}{\text{dim}(r) N^d}
  \frac{\mathrm{d}}{\mathrm{d}m_q}\ln Z_{\text{eff}}^{(r)}
  \sim \frac{1}{\text{dim}(r)}\frac{\mathrm{d}
  f_{\text{mf}}^{(r)}(x,\bar{\lambda}_0)}{\mathrm{d}m_q},
\end{equation}
where $\bar{\lambda}_0$ is the solution of the stationary condition
(see Eq.~(\ref{eq:free_quark})),
\begin{equation}
 \frac{\partial f_{\text{mf}}^{(r)}(x,\bar{\lambda})}
  {\partial\bar{\lambda}}\Biggr|_{\bar{\lambda}=\bar{\lambda}_0}
 =\text{dim}(r)+\sqrt{\frac{d-1}{2}}\cdot\frac{\partial
  f_{\text{mf}}^{(r)}(x,\bar{\lambda})}{\partial m_q}
  \Biggr|_{\bar{\lambda}=\bar{\lambda}_0} =0.
\label{eq:stationary}
\end{equation}
Then, using the stationary condition, we can readily obtain
\begin{align}
 \Psi &=\frac{1}{\text{dim}(r)}\frac{\mathrm{d}f_{\text{mf}}^{(r)}
  (x,\bar{\lambda}_0)}{\mathrm{d}m_q}= \frac{1}{\text{dim}(r)}
  \frac{\partial f_{\text{mf}}^{(r)}(x,\bar{\lambda}_0)}{\partial m_q}
  \notag\\
 &=\frac{1}{\text{dim}(r)}\sqrt{\frac{2}{d-1}}\biggl(\frac{\partial
  f_{\text{mf}}^{(r)}(x,\bar{\lambda}_0)}{\partial\bar{\lambda}_0}
  -\text{dim}(r)\bar{\lambda}_0\biggr)
  =-\sqrt{\frac{2}{d-1}}\bar{\lambda}_0 .
\end{align}
Actually this result does not depend on the concrete form of the free
energy. We can immediately reach the same result from
Eq.~(\ref{eq:identity}) by infinitesimally shifting the integration
variable. Thus, the scalar condensate is simply proportional to the
chiral condensate.

In the same way, we can relate our susceptibility $\chi_\lambda$
defined in Eq.~(\ref{eq:sus_lam}) to the conventional chiral
susceptibility $\chi_\Psi$ which is given by
\begin{align}
 \chi_\Psi &=\frac{1}{[\text{dim}(r)]^2}\frac{\mathrm{d}^2
  f_{\text{mf}}^{(r)}(x,\bar{\lambda}_0)}{\mathrm{d}m_q^2} \notag\\
 &=\frac{1}{[\text{dim}(r)]^2}\Biggl\{\Biggl(\frac{\mathrm{d}
  \bar{\lambda}_0}{\mathrm{d}m_q}\Biggr)^2\frac{\partial^2
  f_{\text{mf}}^{(r)}(x,\bar{\lambda}_0)}{\partial\bar{\lambda}^2}
  +\frac{\partial^2 f_{\text{mf}}^{(r)}(x,\bar{\lambda}_0)}
  {\partial m_q^2}\Biggr\}.
\label{eq:chi_psi}
\end{align}
Differentiating the first equation of Eq.~(\ref{eq:stationary}) by
$\bar{\lambda}$ and $m_q$ respectively and combining them, we can
reach the relation;
\begin{equation}
 \frac{\mathrm{d}\bar{\lambda}_0}{\mathrm{d}m_q}=\text{dim}(r)
  \sqrt{\frac{2}{d-1}}\Biggl(\frac{\partial^2 f_{\text{mf}}^{(r)}(x,
  \bar{\lambda}_0)}{\partial\bar{\lambda}^2}\Biggr)^{-1}
  -\sqrt{\frac{2}{d-1}}.
\end{equation}
In the vicinity of the critical temperature with small current quark
mass ($m_q\sim0$), the curvature of the free energy (second partial
derivative of the free energy) with respect to the scalar meson field
approaches zero. Consequently the first term in the brackets in
Eq.~(\ref{eq:chi_psi}) dominates to result in
\begin{equation}
 \chi_\Psi \sim \frac{2}{d-1}\Biggl(\frac{\partial^2
  f_{\text{mf}}^{(r)}(x,\bar{\lambda}_0)}{\partial\bar{\lambda}^2}
  \Biggr)^{-1} =\frac{2\beta}{d-1}\Bigl(\langle\lambda^2\rangle-
  \langle\lambda\rangle^2\Bigr).
\end{equation}


\section{ZERO TEMPERATURE EXPRESSIONS FOR $N_{\text{f}}$ FLAVORS}
\label{app:zero}

Here we briefly summarize the zero temperature expressions given in
Ref.~\cite{klu83} with modifications for $N_{\text{f}}$ flavors. The
effective potential in terms of the chiral order parameter
$\bar{\lambda}$ is modified at zero temperature into
\begin{equation}
 V_{\text{eff}}(\bar{\lambda})=\frac{N_{\text{c}}}{2}\bar{\lambda}^2
  -\frac{N_{\text{c}}N_{\text{f}}}{4}\ln\biggl[\bar{\lambda}
  +\sqrt{\frac{2}{d}}m_q \biggr],
\end{equation}
from which the stationary value of $\bar{\lambda}$ is obtained as
\begin{equation}
 \bar{\lambda}=-\frac{m_q}{\sqrt{2d}}+\sqrt{\frac{N_{\text{f}}}{4}
  +\frac{m_q^2}{2d}}.
\end{equation}
If we rescale $m_q=\tilde{m}_q\sqrt{N_{\text{f}}/4}$, then the
$N_{\text{f}}$ dependence in $\bar{\lambda}$ can be factorized as
\begin{equation}
 \bar{\lambda}=\sqrt{\frac{N_{\text{f}}}{4}}\Biggl(-\frac{\tilde{m}_q}
  {\sqrt{2d}}+\sqrt{1+\frac{\tilde{m}_q^2}{2d}}\Biggr),
\end{equation}
where the expression in the brackets is nothing but that of
Ref.~\cite{klu83} without any care about the flavor number. Thus we
can expect that most of modifications could be absorbed in the rescale
of the current quark mass $m_q\to\tilde{m}_q$. In fact, the hadron
spectrum for $N_{\text{f}}$ flavors is given by
\begin{align}
 \cosh M_p &= d\Biggl\{\frac{4}{N_{\text{f}}}\Biggl(\frac{m_q}
  {\sqrt{2d}}+\sqrt{\frac{N_{\text{f}}}{4}+\frac{m_q^2}{2d}}\Biggr)^2
  -1\Biggr\}+2p+1 \notag\\
 &= d\Biggl\{\Biggl(\frac{\tilde{m}_q}{\sqrt{2d}}+\sqrt{1
  +\frac{\tilde{m}_q^2}{2d}}\Biggr)^2-1\Biggr\}+2p+1.
\end{align}
Therefore the hadron spectrum necessary for fixing the model
parameters is not affected by the number of flavors.


\section{EXPANSIONS}
\label{app:expansion}

We shall discuss the expansion in terms of $1/\alpha$ for the
integrations $\tilde{I}_n(x;\alpha)$ and $\tilde{I}_n^m(x;\alpha)$
given by Eqs.~(\ref{eq:til_I_1}) and (\ref{eq:til_I_2}). Such
expansions are exploited for the numerical calculation with adjoint
quarks. It is somewhat inefficient to attack the integration given in
Eq.~(\ref{eq:til_I_2}) directly and thus we expand the free energy in
terms of $1/\alpha$. We can obtain such an expansion from
Eqs.~(\ref{eq:til_I_1}) and (\ref{eq:til_I_2}) up to third order as
\begin{align}
 \tilde{I}_n(x;\alpha) &=\frac{1}{2\alpha}\bigl\{I_{n+1}(x)+I_{n-1}(x)
  \bigr\} -\frac{1}{8\alpha^2}\bigl\{I_{n+2}(x)+2I_n(x)+I_{n-2}(x)
  \bigr\} \notag\\
 &\qquad +\frac{1}{24\alpha^3}\bigl\{I_{n+3}(x)+3I_{n+1}(x)
  +3I_{n-1}(x)+I_{n-3}(x)\bigr\},
\label{eq:expan_fund} \\
 \tilde{I}_n^m(x;\alpha) &=\frac{1}{2\alpha}\bigl\{I_{m+1}(x)
  I_{n-1}(x)+I_{m-1}(x)I_{n+1}(x)\bigr\} \notag\\
 &\qquad -\frac{1}{8\alpha^2}\bigl\{I_{m+2}(x)I_{n-2}(x)+2I_m(x)I_n(x)
  +I_{m-2}(x)I_{n+2}(x)\bigr\} \notag\\
 &\qquad +\frac{1}{24\alpha^3}\bigl\{I_{m+3}(x)I_{n-3}(x)
  +3I_{m+1}(x)I_{n-1}(x) \notag\\
 &\qquad\qquad +3I_{m-1}(x)I_{n+1}(x)+I_{m-3}(x)I_{n+3}(x)\bigr\}.
\label{eq:expan_adj}
\end{align}

\begin{figure}
\includegraphics[width=7cm]{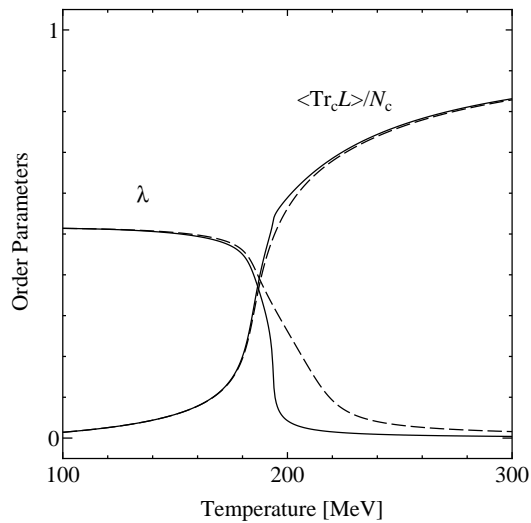}
\caption{Comparison between the numerical results of Ref.~\cite{fuk03}
(shown by dashed curves) and the approximate estimates by the free
energy expanded in terms of $1/\cosh(N_\tau E)$ up to third order
(shown by solid curves).}
\label{fig:expansion}
\end{figure}

We can test the reliability of such approximations by comparing with
the full (not expanded) numerical results. In Fig.~\ref{fig:expansion}
we show the full numerical result obtained from the free energy
(\ref{eq:free_quark}) directly (dashed curve) and the approximate one
with the expanded free energy (solid curve) in the case of fundamental
quarks with Parameter~I. It is obvious from the figure that the
expansion (\ref{eq:expan_fund}) works well except for the behavior of
the chiral order parameter around the chiral phase transition. Thus we
can presume that the expansion (\ref{eq:expan_adj}) will work enough
for adjoint quarks as well, at least, to see the qualitative features.


\end{document}